\documentclass[12pt]{article}
\usepackage{jheppub} % for details on the use of the package, please
\usepackage{float}
\usepackage{amsmath,bbm,blkarray,bm}
\usepackage[vcentermath]{youngtab}
\usepackage{slashed}
\usepackage{booktabs}
\usepackage{multirow}
\usepackage{color}
\usepackage[numbers,sort&compress]{natbib}
\usepackage[T1]{fontenc} % if needed

\usepackage{adjustbox}
\usepackage{cancel}

\def\nn{\nonumber\\ }

\def\entropy{\mathsf{s}}

\renewcommand{\O}{\mathcal{O}}
\newcommand{\op}[3]{\O^{#2,#3}_{#1}}
\newcommand{\opp}[2]{\O^{#2}_{#1}}

\newcommand{\dopp}[2]{\mathcal{P}^{#2}_{#1}}

\newcommand{\Opp}[2]{\mathcal{Q}^{#2}_{#1}}

\newcommand{\Wcc}[2]{C^{#2}_{#1}}
\newcommand{\Wcci}[3]{C^{#3}_{\substack{#1 \\ #2}}}

\allowdisplaybreaks

\title{Dark Matter Effective Field Theory and an Application to Vector Dark Matter}

\author[a]{Jason~Aebischer,}
\author[b]{Wolfgang~Altmannshofer,}
\author[c]{Elizabeth~E.~Jenkins,}
\author[c]{Aneesh V.~Manohar}

\affiliation[a]{Physik-Institut, Universit\"at Z\"urich, CH-8057 Z\"urich, Switzerland}
\affiliation[b]{Department of Physics and Santa Cruz Institute for Particle Physics \\
University of California, Santa Cruz, CA 95064, USA}
\affiliation[c]{Department of Physics, University of California, San Diego, 9500 Gilman Drive,\\ La Jolla, CA 92093-0319, USA}

\emailAdd{jason.aebischer@physik.uzh.ch}
\emailAdd{waltmann@ucsc.edu}
\emailAdd{ejenkins@ucsd.edu}
\emailAdd{amanohar@ucsd.edu}

\abstract{
The Standard Model Effective Field Theory (SMEFT) and the Low Energy Effective Field Theory (LEFT) can be extended by adding additional spin 0, 1/2 and 1 dark matter particles which are singlets under the Standard Model (SM) gauge group. We classify all gauge invariant interactions in the Lagrangian up to terms of dimension six, and present the tree-level matching conditions between the two theories at the electroweak scale. The most widely studied dark matter models, such as those based on the Higgs portal or on kinetic mixing between the photon and a dark photon, are based on dimension-four interactions with the SM sector. We consider a model with dark vector particles with a $\mathbb{Z}_2$ symmetry, so that the lightest dark matter particle is stable. The leading interaction with the SM is through dimension-six operators involving two dark vector field-strength tensors and the electromagnetic field-strength tensor. This model is a viable dark matter model in the freeze-in scenario for a wide range of parameters.
}

\begin{document}
\maketitle

\clearpage

%%%=====================================================================================================================
\section{Introduction}
%%%=====================================================================================================================

Dark Matter (DM) is a viable candidate to explain a number of otherwise unexplained observations in the universe. Evidence for dark matter is astrophysical or cosmological --- galactic rotation curves, gravitational lensing, the cosmic microwave background fluctuations, large scale structure, etc.

Particle physics explanations for DM add one or more new particles which interact weakly with the Standard Model (SM) particles.
To study the implications of additional DM particles in a model-independent way, it is convenient to adopt an Effective Field Theory (EFT) approach. A DM EFT (DMEFT) consists of local operators constructed from SM as well as DM fields. Many DMEFTs already exist in the literature.\footnote{A discussion of non-relativistic DMEFTs can be found in Refs.~\cite{Fan:2010gt,Fitzpatrick:2012ib,Fitzpatrick:2012ix,Bellazzini:2013foa,Cirelli:2013ufw,Catena:2014uqa,Ovanesyan:2014fwa,SuperCDMS:2015lcz,Catena:2019hzw,DelNobile:2021icc,Hoferichter:2015ipa,Hoferichter:2016nvd,Hoferichter:2018acd}, for example.} Various subsets of operators containing DM fields appear in Refs.~\cite{DeSimone:2016fbz,Harnik:2008uu,Kopp:2009qt,Goodman:2010qn,Barman:2020plp,Barman:2020ifq,Cheung:2012gi,Buckley:2013jwa,Fedderke:2014wda,Hisano:2015bma,Bhattacharya:2021edh,Barducci:2021egn}. DM interactions with gluons and quarks as well as their impact on collider searches and direct detection for fermion and/or scalar DM particles are considered in Refs.~\cite{Fox:2011pm,Goodman:2010yf,Goodman:2010ku}. DM fermions coupling to photons are studied in Refs.~\cite{Crivellin:2014gpa,Crivellin:2015wva,Arina:2020mxo}. The authors of Ref.~\cite{Bishara:2016hek} present a DMEFT containing scalars and fermions interacting with quarks, gluons and photons, and study the interactions at low energy scales.
Loop effects from fermion DM particles are described in Ref.~\cite{Crivellin:2014qxa}, and matching conditions for the DMEFT at the EW scale are given in Ref.~\cite{Hill:2014yka}. Furthermore, in the context of co-annihilation, effective DM operators are discussed in Refs.~\cite{Bell:2013wua,Baker:2015qna}.

Several more general analyses of full DMEFTs including higher dimensional operators exist.  In Ref.~\cite{DelNobile:2011uf}, an extension of the SM containing a scalar field in various representations is described. In Ref.~\cite{DeSimone:2013gj}, the SM is extended by a Majorana fermion and a real scalar field, including operators up to dimension eight. A complete basis of operators including the SM fields together with a Majorana fermion is given in Refs.~\cite{Matsumoto:2016hbs,Matsumoto:2014rxa,Han:2017qkr}, with an analysis of the    impact of the operators on astroparticle and collider searches.  A DMEFT which couples scalar, Dirac and vector DM particles to quarks and gluons is presented in Ref.~\cite{Belyaev:2018pqr}. In Ref.~\cite{Duch:2014xda}, a non-redundant set of operators including DM fields of spin $\leq 1$ is considered.   A discrete $\mathbb{Z}_2$ symmetry under which the DM particles are odd, whereas SM particles are even, is imposed so that the DM particles are stable. In Ref.~\cite{Brod:2017bsw}, a general EFT containing Dirac and Majorana fermions as well as complex and real scalar fields up to dimension-seven operators is presented, where the operators are assumed to be invariant under a global $U(1)$ symmetry. In a recent work, the authors of Ref.~\cite{Criado:2021trs} discussed an extension of the SM Effective Field Theory (SMEFT) \cite{Grzadkowski:2010es} with spin 0, 1/2 and 1 particles, presenting a general non-redundant basis of gauge-invariant operators up to dimension six. The DM fields are assumed to transform as electroweak multiplets with arbitrary weak isospin and hypercharge, and to respect a $\mathbb{Z}_2$ symmetry under which DM particles are odd. Recently, a study of portal effective theories (PETs) was presented \cite{Arina:2021nqi}, where electroweak scale PETs encompass all portal operators up to dimension five, while the strong scale PETs additionally contain all portal operators of dimension six and seven that contribute at leading order to quark-flavour violating transitions.

A fully general DMEFT including scalar, fermion and vector DM fields in addition to the SM degrees of freedom, valid above and below the EW scale, is still missing, however. In this work, we close this gap by extending the SMEFT as well as the Low Energy Effective Field Theory (LEFT) \cite{Jenkins:2017jig,Jenkins:2017dyc} by spin 0, 1/2 and 1 DM particles which are singlets under the SM gauge group.   We construct the full set of non-redundant operators involving DM and SM fields up to dimension six, without imposing any underlying symmetry on the DM fields.   The purely SM interactions above the EW scale are given by SMEFT \cite{Grzadkowski:2010es}, and below the EW scale by LEFT \cite{Jenkins:2017jig,Jenkins:2017dyc}, and we do not reproduce these operators here. The operators involving both SM and DM fields above the EW scale form a new EFT, called Dark SMEFT (DSMEFT), which is applicable for DM particles with masses above or below the EW scale, provided the new operators involving DM fields are invariant under the Standard Model $SU(3) \times SU(2) \times U(1)$ gauge symmetry and electroweak symmetry breaking is implemented via the usual Higgs mechanism.

A second EFT involving both SM and DM fields, called Dark LEFT (DLEFT), is a generalization of the LEFT to include spin 0, $1/2$ and $1$ DM singlet particles.  DLEFT is applicable for light DM particles with masses below the EW scale interacting with light SM particles at energies below the EW scale, and does not make any assumptions about $SU(3) \times SU(2) \times U(1)$ invariance. The DLEFT operators are invariant under $SU(3) \times U(1)_{\rm em}$ gauge symmetry.  Like the LEFT, DLEFT does not contain the Higgs boson, and it can be considered without reference to DSMEFT for the light DM and light SM particles interacting at energies below the EW scale.  DLEFT does not assume the Higgs mechanism for $SU(2) \times U(1)$ symmetry breaking.

Operators involving only DM fields are common to both DSMEFT and DLEFT, since only DM singlets are considered.
For theories of light dark matter which use the SM Higgs doublet to spontaneously break electroweak $SU(2) \times U(1)$ gauge symmetry, both DSMEFT and DLEFT apply, and we compute the tree-level matching conditions at the EW scale between the two theories when the heavy SM particles ($t$, $h$, $Z$, $W$) are integrated out, assuming that all dark matter particles are light and not integrated out at the EW scale. The results are easily generalized to the case where some dark matter particles are heavy by dropping operators containing those particles in DLEFT.

In addition to the general operator analysis, we also consider a specific DM matter scenario with vector dark matter which interacts with the SM via dimension-six operators involving three field-strength tensors, $F_{\mu }^{\,\,\nu} X_{a\,\nu }{}^{\alpha} X_{b\,\alpha }{}^{\mu} $ and $\widetilde F_{\mu }^{\,\,\nu} X_{a\,\nu }{}^{\alpha} X_{b\,\alpha }{}^{\mu}$. These are the only interactions of a light dark matter vector particle (i.e.\ in DLEFT) with the SM with dimension $\le 6$, if there is a $\mathbb{Z}_2$ symmetry in the dark sector. We work out the allowed parameter space of this model, and briefly discuss some of the phenomenological implications.

The paper is organized as follows. In Sec.~\ref{sec:EFT}, we describe the assumptions made to construct the DSMEFT and the DLEFT. Sec.~\ref{sec:DMpheno}\ analyzes a model containing two different vector DM particles, interacting with the SM photon through a dimension-six triple-gauge-field-strength interaction.  We study the freeze-out and freeze-in scenarios, and plot the allowed region of parameter space. Conclusions are presented in Sec.~\ref{sec:conclusions}. The operator lists resulting from our general analysis are collected in the appendices.   Appendix~\ref{app:DMops} presents the purely DM operators containing only DM fields up to dimension six.  The purely DM operators in Appendix~\ref{app:DMops} are common to both DSMEFT and DLEFT.  Appendix~\ref{app:DSMEFTops} gives the DSMEFT operators involving both SM and DM fields up to dimension-six operators.  Appendix~\ref{app:DLEFTops} provides the DLEFT operators involving both SM and DM fields up to dimension-six operators.  In all cases, the number of operators is given for $n_\phi$ dark scalars, $n_\chi$ dark fermions, $n_X$ dark gauge bosons, and $n_g$ SM generations.  We also give the tree-level matching conditions between DSMEFT and DLEFT at the electroweak scale.

%%%=====================================================================================================================
\section{Dark Matter EFT}\label{sec:EFT}
%%%=====================================================================================================================

In this section, we describe two new DMEFTs based on the SMEFT and the LEFT, which we call DSMEFT and DLEFT.  The DM fields added to the field content of the effective theories are several copies of the spin 0, 1/2 and 1 fields $\phi$, $\chi$ and $X_\mu$, denoted by the generation index $a$.  The DM fields are assumed to be singlets under the SM gauge group, and in particular to be electrically neutral. The scalar fields $\phi_a$ are real scalar fields, since a complex scalar field can be written in terms of two real scalar fields. The fermion fields $\chi_a$ are right-handed.  We use right-handed DM fermions in order to make our results easier to compare with the extensive literature on right-handed sterile neutrinos.   It is possible to write the same operators in terms of left-handed fermion fields which are the charge-conjugates of the right-handed fermion fields. Majorana fermions can be written in terms of right-handed Weyl fermions. The DM field content used in our analysis is summarized in Table~\ref{tab:DMfields}.

\begin{table}[tbp]
  \centering
  \begin{tabular}{lccc}
  \toprule
    DM fields & $\phi$ & $\chi$ & $X_\mu$
    \\
    \midrule
    Spin & 0 & 1/2 & 1 \\
    Number & $n_\phi$ & $n_\chi$ & $n_X$ \\
    \bottomrule
  \end{tabular}
  \caption{DM fields: $\phi$ is a real scalar field, $\chi$ is a {\it right-handed} fermion, and $X_\mu$ is a vector field. The DM fields are assumed to be singlets under the SM gauge group, so all DM fields are electrically neutral.
  }
  \label{tab:DMfields}
\end{table}

 All DM particles are assumed to be light, with masses smaller than the EW scale, so the DM particles are present in the DMEFTs above as well as below the EW scale. When light $\phi$, $\chi$ and $X_\mu$ fields are added to the SM field content, the most general gauge-invariant Lagrangian up to dimension six operators respecting the Standard Model $SU(3) \times SU(2) \times U(1)$ gauge symmetry defines DSMEFT, which generalizes the SMEFT theory to include light DM gauge particles which are gauge singlets.  When light $\phi$, $\chi$ and $X_\mu$ fields are added to the light SM fields contained in the LEFT, the most general gauge-invariant Lagrangian up to dimension six operators respecting the LEFT gauge symmetry $SU(3) \times U(1)_{\rm em}$
defines DLEFT.  Both DSMEFT and DLEFT contain purely DM operators, operators constructed only using DM fields.  Since all DM particles are gauge singlets, the purely DM operators are identical for DSMEFT and DLEFT.

In order to keep the DSMEFT and DLEFT completely general, we do not impose any stabilizing symmetry on the DM fields.  For specific applications, it is always possible to impose symmetries upon the general set of operators, reducing the operator set to a subset allowed by the symmetries.   Consequently, the results of our operator analysis also apply outside the context of DM, namely for extensions of the SMEFT or LEFT with additional light degrees of freedom which are singlets under the SM gauge interactions.

The complete list of gauge-invariant operators up to mass dimension six built from DM and SM fields are collected in the appendices.  For operators which are not hermitian, we have used $+ \text{h.c.}$ to denote that there are also hermitian conjugate operators which must be included, and which have the complex conjugate coefficient in the Lagrangian. Hermitian operators have a real coefficient in the Lagrangian. The purely DM operators are denoted by $\dopp{}{}$. The operators in DSMEFT are denoted by $\mathcal{Q}$ and those in DLEFT by $\mathcal{O}$, which are the same symbols used for the operators in SMEFT and LEFT, respectively.  If the low-energy dark matter theory DLEFT arises as the low-energy limit of DSMEFT, we can compute the matching between the two dark matter theories. Operator coefficients in DLEFT get an additional matching contribution at the electroweak scale after integrating out the SM top quark, the Higgs boson and the ${W}$ and ${Z}$ gauge bosons. This additional matching piece is shown in the tables in Appendices~\ref{app:DMops} and~\ref{app:DLEFTops}.

The DM sector can have internal or gauge symmetries, which constrains the allowed coefficients. For example, if the $n_\phi$ scalars transform as the fundamental of a $SO(n_\phi)$ symmetry, the $\dopp{\phi^3}{}$ operator is forbidden, and the $\dopp{\phi^4}{}$ operator must be of the form $(\phi_a \phi_a)^2$. If the symmetry is gauged, so that DM particles couple to the $X$ gauge bosons, then ordinary derivatives are replaced by covariant derivatives in the listed operators.

Appendix~\ref{app:DMops} contains the full set of purely DM operators, which are identical for DSMEFT and DLEFT. Appendix~\ref{app:DSMEFTops} gives the DSMEFT operators involving both SM and DM fields up to dimension-six operators. Appendix~\ref{app:DLEFTops} provides the full set of DLEFT operators constructed from DM fields and the light SM fields with masses below the electroweak symmetry breaking scale up to dimension-six operators. In all of the three different operator sets we introduce an arbitrary number of DM particles of each type and count the number of independent operators.  We have checked our results using the Python package \texttt{BasisGen} \cite{Criado:2019ugp}.

In the DM sector, we have included several operators $\dopp{\phi}{} $, $\dopp{\phi^2}{} $, $\dopp{\chi}{} $, $\dopp{\phi}{\text{kin}} $, $\dopp{\chi}{\text{kin}} $, $\dopp{X}{\text{kin}} $ which are usually not considered part of the EFT Lagrangian. The linear term $\dopp{\phi}{} $ can be eliminated by shifting the $\phi$ field. The kinetic terms are usually brought to canonical form proportional to the unit matrix, and the mass terms are usually diagonalized. We have included these operators since there are matching contributions which shift their coefficients, so that even if the operators are put in standard form in DSMEFT, they are no longer in standard form in DLEFT. We would then have to make field transformations to put these operators back in standard form, which affects the matching to all the other operators, resulting in more complicated expressions.

The LEFT is constructed from the SMEFT by integrating out the heavy SM particles ($t$, $h$, $Z$ and $W$) with an EW scale mass, leaving only the light SM particles. The Yukawa couplings of the light fermions are $m_f/v$. For consistency in the EFT power counting, these couplings should formally be treated as order $1/v$, i.e.\ the light fermion Yukawa interactions act formally like dimension-five operators rather than dimension-four operators. This power counting was used in Refs.~\cite{Jenkins:2017jig,Jenkins:2017dyc}  in computing the matching conditions between SMEFT and LEFT. The same analysis applies to the matching conditions between DSMEFT and DLEFT. A DM--SM interaction such as $H^\dagger H \phi^2$ gives a mass to the $\phi$ of order $v^2$ after EW symmetry breaking. Since our $\phi$ field is, \emph{by assumption}, much lighter than the EW scale for it to be included in the DLEFT, the $H^\dagger H \phi^2$ operator has a coefficient suppressed by $m_\phi^2/v^2$, and so the operator is effectively dimension six rather than dimension four. Other suppressed operators are given in Appendix~\ref{app:DLEFTops}.

%%%%%%%%%%%%%%%%%%%%%%%%%%%%%%%%%%%%%%%%%%%%%%
\section{A Dark Sector with Two Vectors}\label{sec:DMpheno}
%%%%%%%%%%%%%%%%%%%%%%%%%%%%%%%%%%%%%%%%%%%%%%

Dark abelian gauge bosons, also known as dark photons, are well studied~\cite{Alexander:2016aln}. They can interact with the SM through the renormalizable kinetic mixing portal (the DSMEFT operator $\mathcal Q_{BX}$ in Table~\ref{tab:DSMEFTdim4})
\begin{equation}
 \mathcal L_\text{mix} = \epsilon_a B_{\mu\nu} X_a^{\mu\nu}~.
 \label{3.1}
\end{equation}
The interaction Eq.~\eqref{3.1} generates mixing between the dark photons $X_a$, and the $U(1)_Y$ SM gauge boson, which is a linear combination of the SM photon and $Z$ boson.  It consequently leads to decays of the dark photons into pairs of SM particles, e.g. $X_a \to e^+ e^-$. For the theory to be viable, the coupling $\epsilon_a$ must be tuned to be much smaller than unity, with typical values $\epsilon_a \sim 10^{-10}$.

In the following, we will consider a different scenario in which the dark sector consists of two dark vector particles, but the kinetic mixing portal operators Eq.~(\ref{3.1}) are not present.\footnote{A similar model but expressed in terms of vector fields instead of field strength tensors has for instance been studied in \cite{Farzan:2012kk,Farzan:2014foo}.} The absence of kinetic mixing with the SM $U(1)_Y$ gauge boson can be ensured by imposing a dark parity symmetry under which the dark vectors are odd $X_a^\mu \to - X_a^\mu$ but all SM particles are even. In such a case, the lightest dark vector is absolutely stable and the only relevant interactions between the dark vectors and  SM particles up to dimension six are the dimension-six DSMEFT operators $\mathcal Q_{BX^2} = B_{\mu}^{\ \, \nu} {X_{a \,\nu}}^{\alpha} {X_{b \,\alpha}}^{\mu}$ and $\mathcal Q_{\widetilde BX^2} = \widetilde B_{\mu}^{\ \,\nu} {X_{a \,\nu}}^{\alpha} {X_{b \,\alpha}}^{\mu}$ from Table~\ref{tab:DSMEFTdim61}.\footnote{  DSMEFT can also have the dimension-six operators $\Opp{HX}{} = (H^\dagger H)  X_{a\,\mu \nu} X_b^{\mu \nu}$ and  $\Opp{H\widetilde X}{} =  (H^\dagger H)\widetilde X_{a\,\mu \nu} X_b^{\mu \nu}$ which respect the dark matter $\mathbb{Z}_2$ symmetry. We neglect these operators --- they do not lead to interactions with SM particles lighter than the electroweak scale at dimension six, as the Higgs coupling to SM particles is $m/v$, and is formally $1/\Lambda$ suppressed in the power counting for light particles. Their coefficients $C_{BX^2 }$ and $C_{\widetilde BX^2} $ do not enter the matching conditions to DLEFT up to dimension six.}  $\mathcal Q_{BX^2}$ and $\mathcal Q_{\widetilde BX^2}$ are antisymmetric in the flavor indices $a,b$, so we need a minimum of two dark vectors for the interaction to exist.

Allowing for dark vector mass terms, we thus consider the Lagrangian
\begin{multline}
 \mathcal L = \mathcal L_\text{SM} - \frac{1}{4} X_1^{\mu\nu} X_{1\,\mu\nu} + \frac{m_1^2}{2} X_1^\mu X_{1\, \mu} - \frac{1}{4} X_2^{\mu\nu} X_{2\,\mu\nu} + \frac{m_2^2}{2} X_2^\mu X_{2\, \mu} \\
 + C_{B X^2} B_{\mu}^{\,\,\nu} {X_{1 \,\nu}}^{\alpha} {X_{2 \, \alpha}}^{\mu} + C_{\widetilde B X^2}  \widetilde B_{\mu}^{\,\,\nu} {X_{1 \, \nu}}^{\alpha} {X_{2 \, \alpha}}^{\mu} ~,
 \label{3.2}
\end{multline}
where the SM $U(1)_Y$ gauge field $B_\mu$ is the linear combination $B_\mu = \cos\theta_W A_\mu - \sin\theta_W Z_\mu$ of the SM photon and the $Z$ boson. The two couplings $C_{B X^2}$ and $C_{\widetilde B X^2}$ have mass dimension $-2$. The dual field strength is $\widetilde B_{\mu\nu} = \frac{1}{2} \epsilon_{\mu\nu\alpha\beta} B^{\alpha \beta}$, where we use the sign convention $\epsilon_{0123}=+1$. The triple field-strength operator only exists if the three fields are \emph{different}, so the interactions must involve two different DM gauge fields $X_1$ and $X_2$.

The vector boson mass terms arise from spontaneous symmetry breaking in the dark sector, e.g.\ by DM scalars.  However, in this work, we do not consider the origin of the DM vector boson masses in detail, and merely assume the mass terms given in the above Lagrangian. A minimal mass generation mechanism is to have a $U(1)$ gauge theory with a complex DM scalar for each $X_\mu$ boson. On spontaneous symmetry breaking, the angular component of the scalar gets eaten to give $X_\mu$ a mass, leaving behind the radial degree of freedom. The additional radial mode does not qualitatively affect the discussion of the model. It is easy to construct scenarios where its coupling to the SM is suppressed by Yukawa couplings or loop factors, and it makes small changes to the expansion rate and relic density of the universe. If the scalar mass is greater than twice the $X_\mu$ mass, any DM scalars produced in the early universe will decay into $X_\mu$ bosons. The scalar and gauge boson masses are independent, since they are proportional to the square-root of the $\phi^4$ coupling, and the gauge coupling, respectively.

$\Lambda$ is the new physics scale where one expects new degrees of freedom that interact with both the SM and the dark sector. Interactions with these new degrees of freedom can generate the dimension-six interactions in Eq.~\eqref{3.2} with $C_{B X^2}$ and $C_{\widetilde B X^2}$ proportional to $1/\Lambda^2$.  If the new physics is weakly coupled, the dimension-six operators are generated at one-loop and have a $1/(16\pi^2)$ suppression. However strongly-interacting new physics theories need not have this suppression~\cite{Manohar:2013rga}. In our plots, we use $C_{B X^2} = 1/\Lambda^2$ as the definition of the new physics scale $\Lambda$. Without loss of generality we assume that $X_1$ is the lighter of the two dark vectors, $m_1 < m_2$, and therefore an absolutely stable dark matter candidate.

The vector $X_2$ is not stable; it can decay into $X_1$ and a photon, and, if kinematically allowed, into $X_1$ and a $Z$ boson. Three-body decays mediated by a virtual photon or $Z$ boson are necessarily suppressed compared to the two-body decay $X_2 \to X_1 \gamma$, and are neglected in the following. For the decay rates $X_2 \to X_1 \gamma$ and $X_2 \to X_1 Z$ we find
\begin{equation} \label{eq:X2X1ga}
 \Gamma(X_2 \to X_1 \gamma) = \frac{\cos^2\theta_W}{96 \pi} \big(C^2_{B X^2} + C^2_{\widetilde B X^2} \big) m_2^5  (1+\varrho)(1-\varrho)^3 ~,
\end{equation}
\begin{multline} \label{eq:X2X1Z}
 \Gamma(X_2 \to X_1 Z) = \frac{\sin^2\theta_W}{96 \pi} m_2^5\ \lambda^\frac{1}{2}(1,\varrho,z) \Big[ C^2_{B X^2} \big(1+\varrho + z\big) \lambda(1,\varrho,z)  \\
+  C^2_{\widetilde B X^2} \big( 1 - (\varrho+z) - (\varrho^2+z^2) + (\varrho^3+z^3) +(6 - \varrho - z)\varrho z \big) \Big] ~,
\end{multline}
where $\varrho = m_1^2/m_2^2$ is the ratio of squared dark vector masses, $z = m_Z^2/m_2^2$, and
\begin{align}
\lambda(a,b,c) \equiv a^2+b^2+c^2-2(ab+ac+bc)\,.
\label{lambda}
\end{align}
We are mainly interested in the regime where the $X_2 \to X_1+Z$ decay is kinematically forbidden.

If $X_2$ is produced in the early Universe, there are two viable regimes of $X_2$ lifetime $\tau_{X_2}$, where $\tau_{X_2}^{-1} = \Gamma(X_2 \to X_1\gamma) + \Gamma(X_2 \to X_1 Z)$. Either $X_2$ decays sufficiently quickly so that its decay does not significantly disrupt the element abundances predicted by big bang nucleosynthesis (BBN), or $X_2$ has an extremely long lifetime such that it can be effectively treated as stable, and the observed dark matter abundance has both a $X_1$ and a $X_2$ component. For a quickly decaying $X_2$,  the BBN constraints are typically in the range $\tau_{X_2} \lesssim 1$\,s to $\tau_{X_2} \lesssim 10^4$\,s, depending on the type and the energy spectrum of the visible decay products of $X_2$~\cite{Ellis:1990nb, Cyburt:2002uv, Kawasaki:2017bqm, Dienes:2018yoq}.  We will use the conservative limit  $\tau_{X_2} \lesssim 1$\,s. A late-decaying $X_2$ component of dark matter can lead to distortions of the cosmic microwave background (CMB). If $X_2$ constitutes a significant fraction of the dark matter, its lifetime needs to be larger than approximately $\tau_{X_2} \gtrsim 10^{26}$\,s~\cite{Hu:1993gc, Slatyer:2016qyl, Poulin:2016anj, Dienes:2018yoq}. The two constraints combine to exclude the region $1\, \text{s} \le \tau_{X_2} \le 10^{26} \, \text{s}$.

If both $X_1$ and $X_2$ are sufficiently light, there is an additional contribution to the invisible $Z$ width
\begin{multline} \label{eq:ZX1X2}
 \Gamma(Z \to X_1 X_2) = \frac{\sin^2\theta_W}{96 \pi} m_Z^5 \lambda^\frac{1}{2}(1,x_1,x_2) \Big[ C^2_{B X^2} \big(1+x_1+x_2\big) \lambda(1,x_1,x_2)  \\
 +  C^2_{\widetilde B X^2} \big( 1 - (x_1+x_2) - (x_1^2+x_2^2) + (x_1^3+x_2^3) +(6-x_1-x_2)x_1x_2 \big) \Big] ~,
\end{multline}
with $x_1 = m_1^2/m_Z^2$ and $x_2 = m_2^2/m_Z^2$. LEP measurements of the invisible width of the $Z$ boson imply $\Gamma(Z \to X_1 X_2) < 2.0$\,MeV at 95\% C.L.~\cite{ALEPH:2005ab}.

\subsection{Dark Matter Production}

Having defined the model, we consider in the following two possibilities for dark matter production: freeze-out and freeze-in. In both cases we discuss which values of the new physics parameters (the masses $m_1$ and $m_2$ and new physics scale $\Lambda$) can give the observed dark matter abundance $\Omega h^2 \simeq 0.12$~\cite{Planck:2018vyg}.

\subsubsection{Freeze-out}

%%%%%%%%%%%%%%%%%%%%%%%%%%%%%%%%%
\begin{figure}[tb]
\centering
\includegraphics[width=4cm]{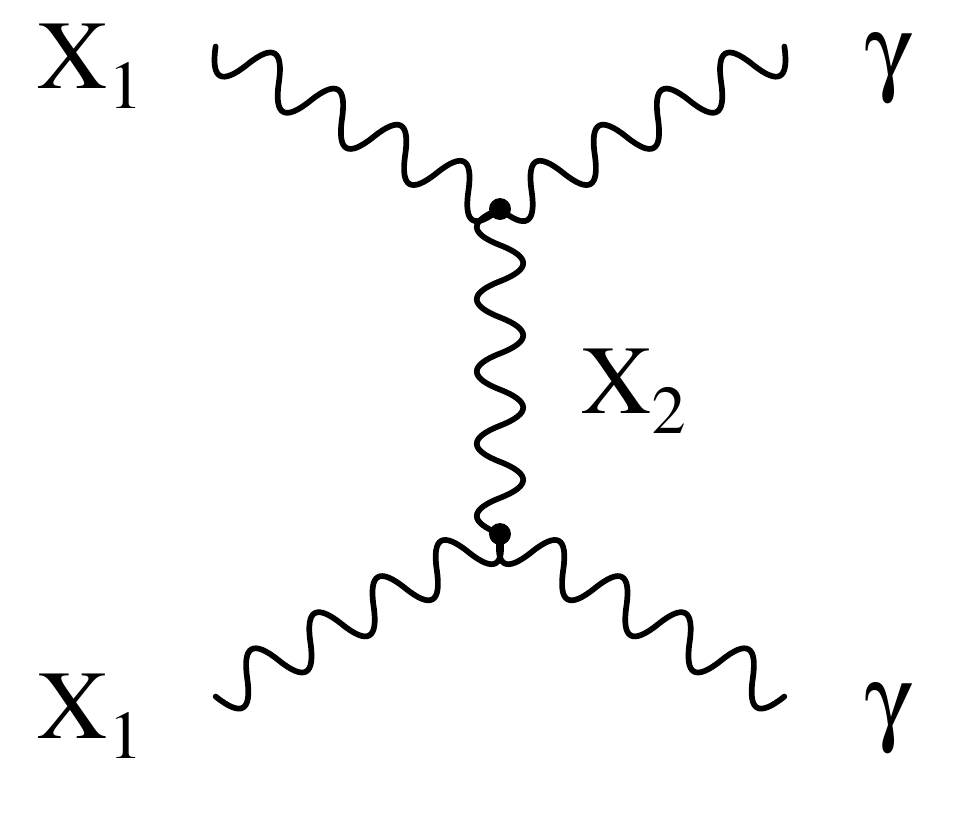} \hspace{2cm}
\raisebox{0.1cm}{\includegraphics[width=4cm]{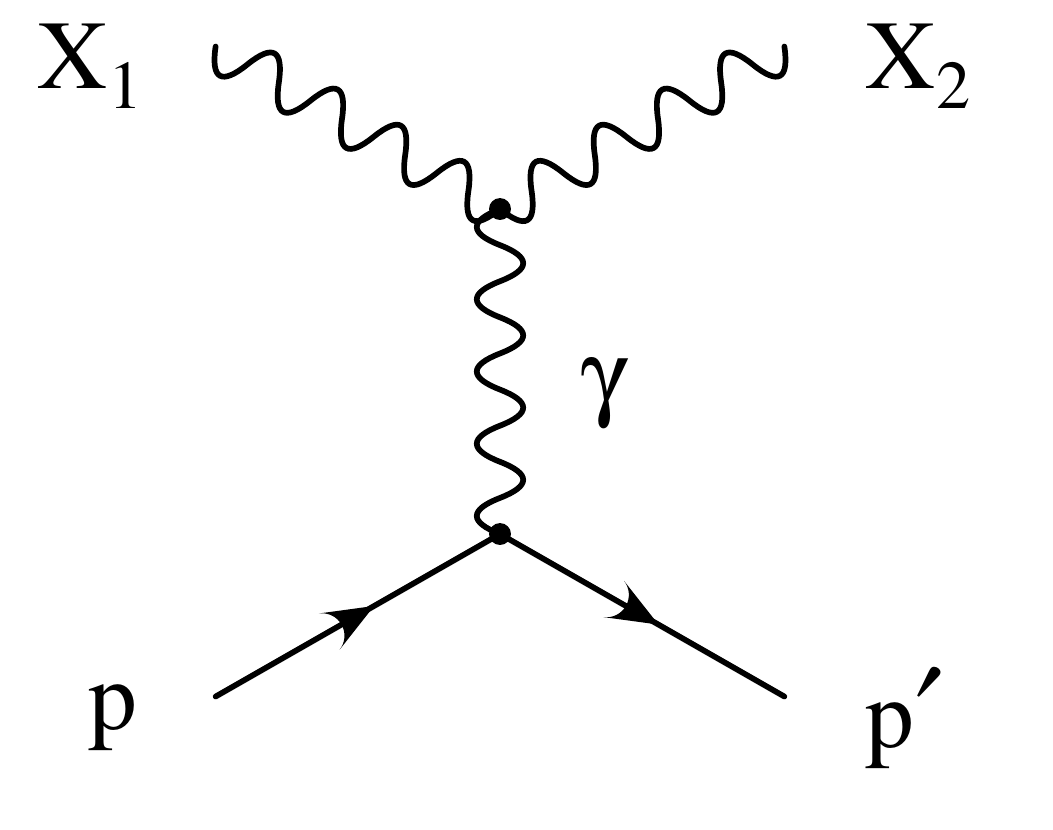}}
\caption{Left: Example $t$-channel diagram for dark matter annihilation into photons. There is also the crossed $u$-channel diagram. Right: Diagram for dark matter scattering off nuclei.}
\label{fig:diagrams}
\end{figure}
%%%%%%%%%%%%%%%%%%%%%%%%%%%%%%%%%%

If the interactions between the SM sector and the dark sector are large enough that the two sectors are in thermal equilibrium in the early universe, the dark matter abundance is set by freeze-out as the universe cools. The correct dark matter relic abundance can be obtained if the annihilation cross section of dark matter into SM particles is of electroweak size. To a good approximation one finds
\begin{eqnarray} \label{eq:freeze-out}
 \Omega h^2 \simeq 0.12 \times \left( \frac{2.2 \times 10^{-26} \text{cm}^3/\text{s}}{\langle \sigma v_\text{rel} \rangle } \right)~,
\end{eqnarray}
where $\langle \sigma v_\text{rel} \rangle$ is the thermal average of the dark matter annihilation cross section times the relative dark matter velocity.

Dark matter annihilation can occur through several channels, in particular $X_1 X_1 \to \gamma \gamma$, $X_1 X_1 \to Z \gamma$, and $X_1 X_1 \to ZZ$ through $t$-channel or $u$-channel $X_2$ exchange, see the left diagram in Figure~\ref{fig:diagrams}. Co-annihilation channels like $X_1 X_2 \to W W$, $X_1 X_2 \to Z h$, or $X_1 X_2 \to \bar f f$ through an $s$-channel photon or $Z$ boson, where $f$ is a SM fermion, can be relevant for $m_1 \simeq m_2$, so that there is an appreciable density of $X_2$. For the $X_1X_1 \to \gamma \gamma$ annihilation cross section, we find
\begin{multline} \label{eq:X1X1gaga}
 \sigma(X_1 X_1 \to \gamma \gamma) v_\text{rel} = \cos^4\theta_W \big(C^2_{B X^2} + C^2_{\widetilde B X^2} \big)^2\ \frac{m_1^6}{144 \pi}\
  \biggl\{  \frac{5\varrho^2}{(1+\varrho)^2} \\ + v_\text{rel}^2   \frac{\varrho^2(167 + 194 \varrho + 71 \varrho^2)}{12(1+\varrho)^4} \biggr\} + O(v_\text{rel}^4) ~,
\end{multline}
where as above, $\varrho = m_1^2/m_2^2$, and we  have expanded to second order in the relative velocity. An analogous expression holds for the cross section $\sigma(X_2 X_2 \to \gamma \gamma) v_\text{rel}$, with $m_1 \leftrightarrow m_2$.

The thermal averages of powers of the relative velocity scale as $\langle v_\text{rel}^{2n} \rangle \sim x^{-n}$, where $x^{-1} = T_\text{fo}/m$ with $m$ the dark matter mass and $T_\text{fo}$ the temperature at freeze-out. Typically, $T_\text{fo}/m \sim 1/20$. Therefore, it is a good approximation to keep only the leading term in the $v_\text{rel}$ expansion. Based on the $X_1 X_1 \to \gamma \gamma$ annihilation channel alone, we find
\begin{equation}
 \langle \sigma v_\text{rel} \rangle = \frac{5 \cos^4\theta_W}{144 \pi} \big(C^2_{BX^2} + C^2_{\widetilde BX^2} \big)^2 m_1^6 \frac{\varrho^2}{(1+\varrho)^2} + \mathcal O(T_\text{fo}/m) ~.
\end{equation}
Using the approximate expression for the relic abundance from above, and demanding that $X_1$ gives all of the dark matter, we find a new physics scale of
\begin{align}
 \Lambda \sim \big(C^2_{BX^2} + C^2_{\widetilde B X^2} \big)^{-\frac{1}{4}} \sim 104\,\text{GeV} \times \left(\frac{m_1}{50\, \text{GeV}}\right)^\frac{3}{4}   \left[ \frac{4 \varrho^2}{(1+\varrho)^2} \right]^\frac{1}{8} ~.
\end{align}
This scale is close to the electroweak scale for $m_1 \sim 50$\,GeV, and gets even smaller for lighter $X_1$. Including the other annihilation channels will modify this estimate for $\Lambda$ by an amount of order one. Such a low new physics scale is challenged by the absence of direct evidence for new degrees of freedom at the LHC and we therefore will not pursue the freeze-out scenario in more detail.

\subsubsection{Freeze-in}

In the freeze-in scenario, the interactions between the SM and the dark sector are so feeble that the two sectors do not reach thermal equilibrium. We assume that after reheating there is a negligible abundance of dark matter, and dark matter particles get produced from decays or scattering of SM particles~\cite{Hall:2009bx}. In our setup, higher dimensional operators parameterize the interactions between the dark matter sector and the SM, similar to the scenarios discussed in Ref.~\cite{Elahi:2014fsa}. For a sufficiently large new physics scale $\Lambda$, the interactions are feeble enough to realize the dark matter freeze-in scenario. The formulation in terms of higher dimensional operators is consistent as long as the reheat temperature $T_\text{rh}$ is much smaller than the new physics scale $\Lambda$.

We briefly review the freeze-in formalism, following largely Ref.~\cite{DEramo:2017ecx}.
The number densities $n_{X_1}$ and $n_{X_2}$ of the dark sector particles $X_1$ and $X_2$ are determined by the Boltzmann equations
\begin{multline}
 \dot n_{X_1} + 3 H n_{X_1} = C_{\gamma\gamma \to X_1 X_1} + C_{Z\gamma \to X_1 X_1} + C_{ZZ \to X_1 X_1} \\*
 + C_{Zh \to X_1 X_2} + C_{WW \to X_1 X_2} + \sum_f C_{\bar f f \to X_1 X_2} ~,
\end{multline}
\begin{multline}
 \dot n_{X_2} + 3 H n_{X_2} = C_{\gamma\gamma \to X_2 X_2} + C_{Z\gamma \to X_2 X_2} + C_{ZZ \to X_2 X_2} \\*
 + C_{Zh \to X_1 X_2} + C_{WW \to X_1 X_2} + \sum_f C_{\bar f f \to X_1 X_2}~,
\end{multline}
where $H$ is the Hubble parameter, and the collision terms on the right-hand side take into account dark matter production from the scattering processes $\gamma\gamma \to X_i X_i$, $Z\gamma \to X_i X_i$, and $ZZ \to X_i X_i$ that involve two dimension-six interactions, as well as the processes $Zh \to X_1 X_2$, $WW \to X_1 X_2$, and $\bar f f \to X_1 X_2$ that are mediated by an $s$-channel $Z$ or $\gamma$ and involve only a single dimension-six interaction. The collision terms are given explicitly in Eqs.~\eqref{3.19}--\eqref{3.24}, and include the final state multiplicities. The $X_1 X_2$ processes, which are order $1/\Lambda^4$, dominate over the $X_1 X_1$ and $X_2 X_2$ processes of order $1/\Lambda^8$ if $m_1 \sim m_2$. Otherwise, the relative rates depend also on the Boltzmann suppression for $X_2$.

It is convenient to rewrite the Boltzmann equations in terms of the comoving number densities $Y_{X_i} = n_{X_i}/\entropy$ where $\entropy$ is the entropy density. One finds
\begin{multline} \label{eq:X1production}
 Y_{X_1} = \int_{T_0}^{T_\text{rh}} \frac{dT}{T} \frac{1}{\entropy H} \big( C_{\gamma\gamma \to X_1 X_1} + C_{Z\gamma \to X_1 X_1} + C_{ZZ \to X_1 X_1} \\
 + C_{Zh \to X_1 X_2} + C_{WW \to X_1 X_2} + \sum_f C_{\bar f f \to X_1 X_2} \big) ~,
\end{multline}
\begin{multline} \label{eq:X2production}
Y_{X_2} = \int_{T_0}^{T_\text{rh}} \frac{dT}{T} \frac{1}{\entropy H} \big( C_{\gamma\gamma \to X_2 X_2} + C_{Z\gamma \to X_2 X_2} + C_{ZZ \to X_2 X_2} \\
 + C_{Zh \to X_1 X_2} + C_{WW \to X_1 X_2} + \sum_f C_{\bar f f \to X_1 X_2} \big) ~,
\end{multline}
where $T_0 \simeq 0$ is the temperature today, and the entropy density and  Hubble parameter are given by
\begin{align}
 \entropy &= \frac{2 \pi^2}{45} g_\star T^3 ~, & H &= \frac{\pi}{3 \sqrt{10}} \sqrt{g_\star} \frac{T^2}{M_\text{Pl}}~,
\end{align}
with $M_\text{Pl} \simeq 2.4 \times 10^{18}$\,GeV the reduced Planck mass, and $g_\star$  the effective number of degrees of freedom. In Eqs.~\eqref{eq:X1production}, \eqref{eq:X2production}, we have approximated the effective number of degrees of freedom in the thermal bath for the entropy and energy density as a constant, $g_\star^s = g_\star^\rho = g_\star = \text{const}$. This is a good approximation for temperatures above the electroweak scale~\cite{Husdal:2016haj}. For such temperatures, $g_\star \simeq 427/4$ in the SM.

We identify two qualitatively different scenarios that are determined by the hierarchy of the scales $m_1$, $m_2$, and $T_\text{rh}$:
\begin{itemize}
 \item[(1)] $m_1 < T_\text{rh} < m_2$: In this case only $X_1$ can be produced by the scattering of SM particles. The $s$-channel production modes that include $X_2$ in the final state are kinematically suppressed. The dark matter abundance is therefore given by
\begin{equation}
 \Omega h^2 = \frac{h^2 \entropy_0}{\rho_\text{crit}}  m_1 Y_{X_1} \simeq (2.7 \times 10^8) \ \left( \frac{m_1}{1\,\text{GeV}}\right) Y_{X_1} ~,
\end{equation}
where $\rho_\text{crit} = 1.053 672(24) \times 10^{-5} h^2\,\text{GeV}/\text{cm}^{3}$~\cite{ParticleDataGroup:2020ssz} is the critical density of the Universe and $\entropy_0 = 2891.2(1.9) /\text{cm}^3$~\cite{ParticleDataGroup:2020ssz} is the entropy density today.
 If the reheat temperature $T_\text{rh}$ is significantly below the electroweak scale, the only relevant process that produces dark matter, and that needs to be taken into account in the calculation of comoving number density $Y_{X_1}$ is $\gamma\gamma \to X_1 X_1$. For $T_\text{rh}$ above the electroweak scale, $Z\gamma \to X_1 X_1$ and $ZZ \to X_1 X_1$ also need to be considered.
 \item[(2)] $m_1 < m_2 < T_\text{rh}$: In this case, both $X_1$ and $X_2$ are produced by the scattering of SM particles. To obtain the dark matter abundance we need to distinguish two sub-cases:
 \begin{itemize}
 \item[(2a)] the lifetime of $X_2$ is much larger than the age of the universe. In this case dark matter is made from two components, the absolutely stable $X_1$ and the approximately stable $X_2$ and the corresponding abundances add up
 \begin{equation}
 \Omega h^2 = \frac{h^2 \entropy_0}{\rho_\text{crit}} \big( m_1 Y_{X_1} + m_2 Y_{X_2} \big) \simeq (2.7 \times 10^8)\ \left[ \left( \frac{m_1}{1\, \text{GeV}}\right) Y_{X_1} + \left( \frac{m_2}{1\,\text{GeV}} \right) Y_{X_2} \right] ~.
 \end{equation}
 \item[(2b)]the lifetime of $X_2$ is much shorter than the age of the universe. In this case, $X_2$ has decayed through the processes $X_2 \to X_1 \gamma$ and, if kinematically allowed, $X_2 \to X_1 Z$, producing one $X_1$ particle per decaying $X_2$ particle. The relic density therefore is
 \begin{equation}
 \Omega h^2 = \frac{h^2 \entropy_0}{\rho_\text{crit}} m_1 \big( Y_{X_1} + Y_{X_2} \big) \simeq (2.7 \times 10^8)\ \left( \frac{m_1}{1\,\text{GeV}} \right) \big( Y_{X_1} + Y_{X_2} \big) ~.
 \end{equation}
 \end{itemize}
For a reheat temperature below the electroweak scale, only processes with photons and light SM fermions need to be taken into account when calculating $Y_{X_1}$ and $Y_{X_2}$. For sufficiently large $T_\text{rh}$, all processes in~\eqref{eq:X1production} and \eqref{eq:X2production} are relevant.
\end{itemize}

To evaluate the relic density, we need to determine the various collision terms. In the case of $2 \to 2$ scattering one has the following generic expression
\begin{multline} \label{eq:C_abcd}
 C_{ab \to cd} \simeq \int \frac{d^3 p_a}{(2\pi)^3 2 E_a} \int \frac{d^3 p_b}{(2\pi)^3 2 E_b} \int \frac{d^3 p_c}{(2\pi)^3 2 E_c} \int \frac{d^3 p_d}{(2\pi)^3 2 E_d} \\
 \times f_a^\text{eq}(E_a) f_b^\text{eq}(E_b) |\overline{\mathcal M}_{ab\to cd}|^2 (2\pi)^4 \delta^{(4)}(p_a + p_b - p_c - p_d) ~,
\end{multline}
since the final state occupation numbers $f_c^\text{eq}(E_c),\ f_d^\text{eq}(E_d) \ll 1$ so we can neglect final state Bose and Fermi factors.
Note that the squared matrix elements are summed over both initial and final state degrees of freedom (e.g. spins, polarizations, color). The equilibrium phase space distributions are to a good approximation given by $f_i^\text{eq}(E_i) \simeq \exp(- E_i/T)$. Carrying out the momentum integrations, the collision terms for the scattering processes can be related to the corresponding scattering cross sections. We find (including the multiplicity of the final state in the definition of the collision integral)
\begin{align}
 C_{\gamma \gamma \to X_i X_i} &= \frac{1}{8\pi^4} T \int_0^\infty ds~  s^\frac{3}{2} K_1(\sqrt{s}/T) \sigma(\gamma \gamma \to X_i X_i) ~,
 \label{3.19} \\
 C_{ZZ \to X_i X_i} &= \frac{9}{32\pi^4} T \int_{4m_Z^2}^\infty ds~  s^\frac{3}{2} K_1(\sqrt{s}/T) \left(1-\frac{4m_Z^2}{s}\right) \sigma(ZZ \to X_i X_i) ~, \\
 C_{Z \gamma \to X_i X_i} &= \frac{3}{8\pi^4} T \int_{m_Z^2}^\infty ds~  s^\frac{3}{2} K_1(\sqrt{s}/T) \left(1 - \frac{m_Z^2}{s} \right)^2 \sigma(Z \gamma \to X_i X_i) ~, \\
 C_{WW \to X_1 X_2} &= \frac{9}{32\pi^4} T \int_{4m_W^2}^\infty ds~  s^\frac{3}{2} K_1(\sqrt{s}/T) \left(1 - \frac{4m_W^2}{s} \right) \sigma(WW \to X_1 X_2)~, \\
 C_{Zh \to X_1 X_2} &= \frac{3}{32\pi^4} T \int_{(m_Z+m_h)^2}^\infty \hspace{-0.25cm} ds~  s^\frac{3}{2} K_1(\sqrt{s}/T)\, \lambda\!\left(1,\frac{m_Z^2}{s},\frac{m_h^2}{s}\right) \sigma(Zh \to X_1 X_2)~, \\
 C_{\bar f f \to X_1 X_2} &= \frac{N_c^2}{8\pi^4} T \int_{4m_f^2}^\infty ds~  s^\frac{3}{2} K_1(\sqrt{s}/T) \left(1 - \frac{4m_f^2}{s} \right) \sigma(\bar f f \to X_1 X_2)~,
 \label{3.24}
\end{align}
with a color factor $N_c = 3$ for quarks and $N_c = 1$ for leptons.
$\lambda(a,b,c)$ is defined in Eq.~\eqref{lambda}, and $K_1$ is the first modified Bessel function of the second kind.

In the following we will focus on two benchmark cases that are representative for the generic scenarios (1) and (2) identified above. A comprehensive discussion of the entire parameter space is beyond the scope of this work.

For the first benchmark case we assume that the reheat temperature is below both the electroweak scale and the mass of the second vector $X_2$. Thus, only the $\gamma \gamma \to X_1 X_1$ collision term is relevant. Furthermore, if the mass of $X_1$ is sufficiently small, $m_1 \ll T_\text{rh} \ll m_2$, we find the following simple expression for the cross section
\begin{equation}
 \sigma(\gamma \gamma \to X_1 X_1) \simeq  \frac{263 \cos^4\theta_W}{215040 \pi} \frac{s^5}{m_2^4} \big( C_{BX^2}^2 + C_{\widetilde B X^2}^2 \big)^2 ~,
 \label{3.26}
\end{equation}
where $s$ is the diphoton invariant mass.
Neglecting the temperature dependence of the effective number of degrees of freedom $g_\star$, the corresponding collision term and the resulting comoving $X_1$ number density can be determined analytically. We find
\begin{equation}
 C_{\gamma \gamma \to X_1 X_1} \simeq \frac{9089280 \cos^4\theta_W}{\pi^5} \frac{T^{16}}{m_2^4 } \big( C_{BX^2}^2 + C_{\widetilde B X^2}^2 \big)^2 ~,
\end{equation}
\begin{equation}
 Y_{X_1} \simeq \frac{613526400 \sqrt{10}}{11 \pi^8} \frac{\cos^4\theta_W}{g_\star^{3/2}} \frac{M_\text{pl} T_\text{rh}^{11}}{m_2^4}\big( C_{BX^2}^2 + C_{\widetilde B X^2}^2 \big)^2~.
\end{equation}
If the reheat temperature is above the electroweak scale, the comoving number density increases by an $\mathcal O(1)$ amount due to the additional $Z Z \to X_1 X_1$ and $\gamma Z \to X_1 X_1$ channels, but the characteristic dependence on $T_\text{rh}^{11}$ does not change. The correct dark matter abundance, $\Omega h^2 \simeq 0.12$, is obtained for a new physics scale $\Lambda$ approximately 3 orders of magnitude above the reheat temperature
\begin{equation}
\frac{\Lambda}{T_\text{rh}} =
\frac{ \big( C_{BX^2}^2 + C_{\widetilde B X^2}^2 \big)^{-\frac{1}{4}} }{T_\text{rh}} \sim 700 \times \left(\frac{m_1}{1\,\text{GeV}}\right)^\frac{1}{8}\times \left(\frac{T_\text{rh}}{100\,\text{GeV}}\right)^\frac{3}{8} \times \left(\frac{1\,\text{TeV}}{m_2}\right)^\frac{1}{2} \times \left(\frac{100}{g_\star}\right)^\frac{3}{16} .
\label{3.29}
\end{equation}
This is consistent with our assumption that the EFT description of the dark vector interactions is appropriate to determine the dark matter abundance.

In the second benchmark case we assume again that the reheat temperature is below the electroweak scale but this time both $X_1$ and $X_2$ are significantly lighter, $m_1, m_2 \ll T_\text{rh}$. In this case, the dominant dark matter production process is $f \bar f \to X_1 X_2$ mediated by an $s$-channel photon. The corresponding amplitude is suppressed by only one power of the dimension-six dark vector interaction with the photon.
For the cross section we find (assuming for simplicity also $m_f \ll T_\text{rh}$)
\begin{equation}
 \sigma(\bar f f \to X_1 X_2) \simeq \frac{Q_f^2 e^2 \cos^2\theta_W}{96 \pi N_c} s \big( C_{BX^2}^2 + C_{\widetilde B X^2}^2 \big) ~,
\end{equation}
where $Q_f$ is the electric charge of the fermion $f$. The corresponding collision term and the resulting comoving number densities are
\begin{equation}
 C_{\bar f f \to X_1 X_2} \simeq N_c Q_f^2 \, \frac{e^2 \cos^2\theta_W}{\pi^5}  T^{8} \big( C_{BX^2}^2 + C_{\widetilde B X^2}^2 \big) ~,
 \label{3.31}
\end{equation}
\begin{equation}
 Y_{X_1} \simeq Y_{X_2} \simeq \frac{45 \sqrt{10}}{2 \pi^8} \sum_f N_c Q_f^2 \, \frac{e^2 \cos^2\theta_W}{g_\star^{3/2}} {M_\text{pl} T_\text{rh}^{3}}\big( C_{BX^2}^2 + C_{\widetilde B X^2}^2 \big)~.
\end{equation}
For reheat temperatures above the electroweak scale, the comoving number densities will be larger by an $\mathcal O(1)$ amount, because additional dark matter production channels open up. The dominant channels all scale with $T_\text{rh}^3$. Summing over all charged leptons and the five light quarks, $\sum_f N_c Q_f^2 = 20/3$. Assuming the vector $X_2$ decays sufficiently fast, we find the following new physics scale to reproduce the observed dark matter abundance
\begin{equation}
\frac{\Lambda}{T_\text{rh}} = \frac{\big( C_{BX^2}^2 + C_{\widetilde B X^2}^2 \big)^{-\frac{1}{4}}}{T_\text{rh}} \sim   \left( 1.43 \times 10^5 \right) \times \left(\frac{m_1}{1\,\text{GeV}}\right)^\frac{1}{4}\times \left(\frac{100\,\text{GeV}}{T_\text{rh}}\right)^\frac{1}{4} \times \left(\frac{100}{g_\star}\right)^\frac{3}{8} ~.
\label{3.33}
\end{equation}

%%%%%%%%%%%%%%%%%%%%%%%%%%%%%%%%%
\begin{figure}[tb]
\centering
\includegraphics[height=0.48\textwidth]{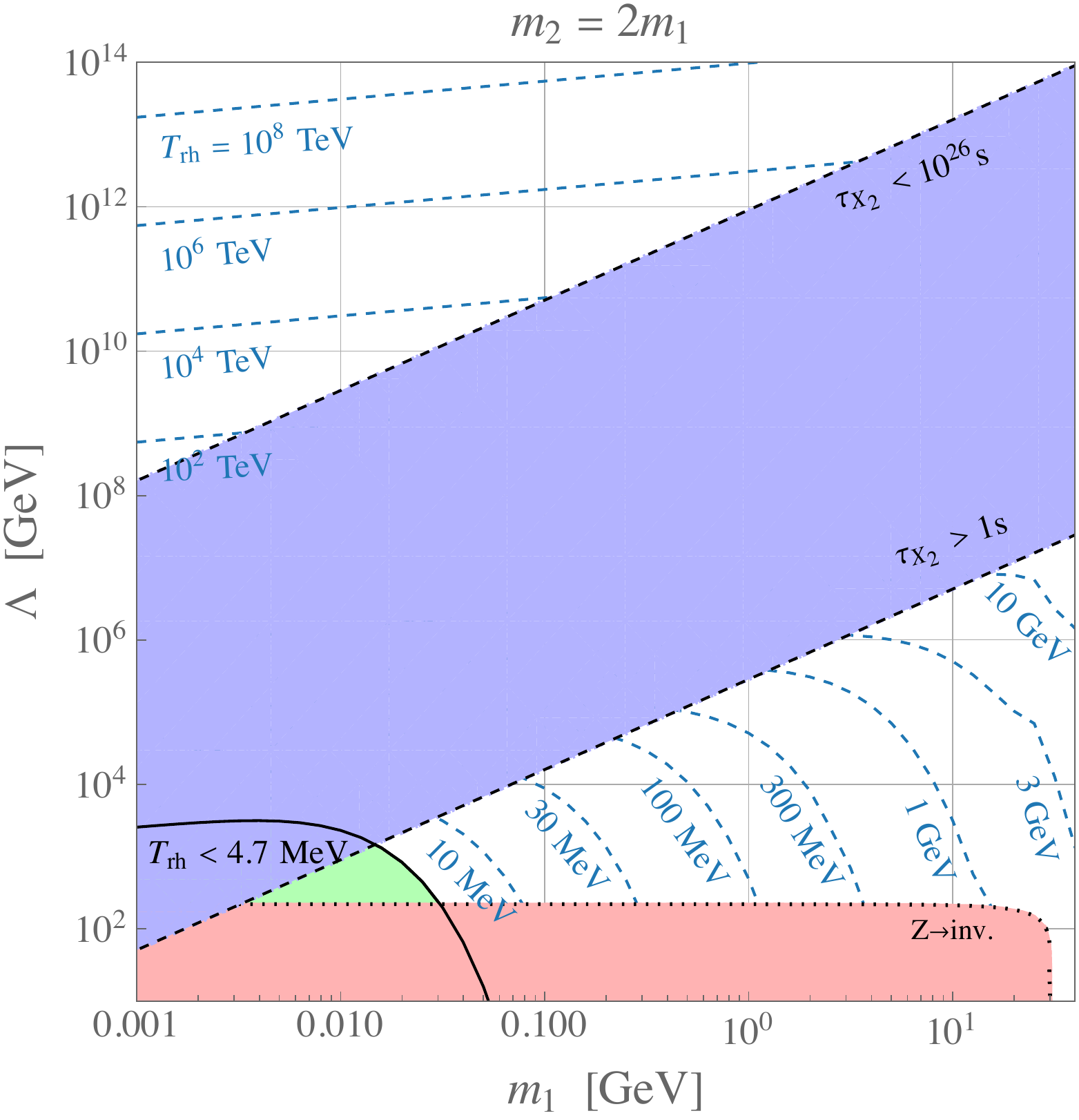}\hspace{1cm}
\includegraphics[height=0.48\textwidth]{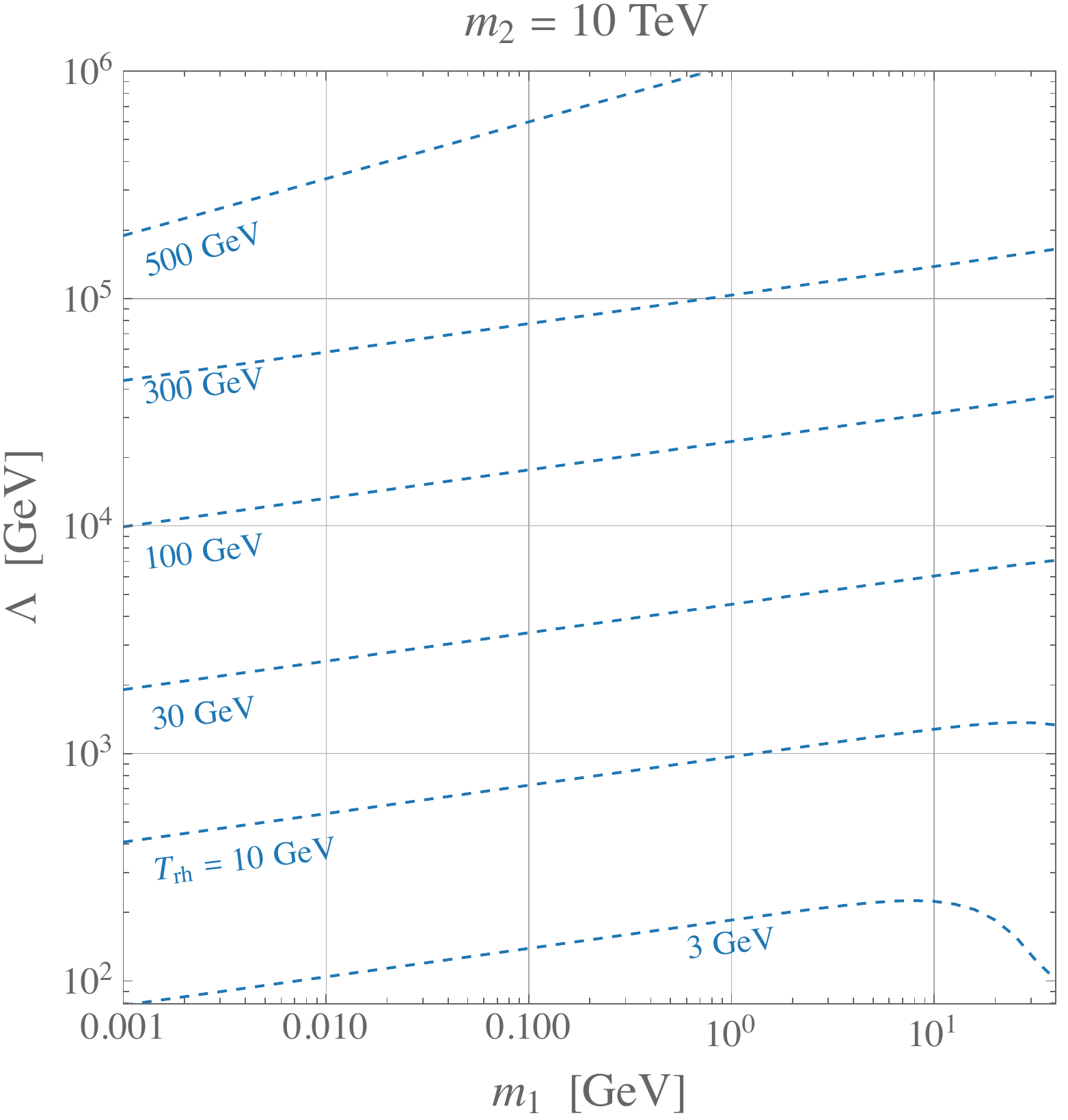}

\caption{Reheat temperature to reproduce the observed dark matter abundance. The curves have been evaluated for $C_{\widetilde B X^2}=0$ and $C_{B X^2}=1/\Lambda^2$. The red shaded region is excluded by the invisible width of the $Z$, the blue shaded region by lifetime constraints on $X_2$, and the green shaded region by the limit $T_{\text{rh}} < 4.7$~MeV from the CMB and BBN~\cite{deSalas:2015glj}.}
\label{fig:Trh_plots}
\end{figure}
%%%%%%%%%%%%%%%%%%%%%%%%%%%%%%%%%%

In Fig.~\ref{fig:Trh_plots}, we show the new physics scale $\Lambda$ as a function of $m_1$, the mass of the dark matter particle, for different values of the reheat temperature $T_{\text{rh}}$. In both plots, the new physics scale $\Lambda$ is always considerably greater than $T_{\text{rh}}$ so the EFT computation of the cross sections is valid. For simplicity, we only show the case where $C_{\widetilde B XX}=0$. Including both the $B XX $ and $\widetilde BXX$ operators makes only small differences in the plots. In the plots, we have included the full mass-dependence of the cross sections, rather than the simplified expressions in Eqs.~\eqref{3.26}, \eqref{3.31}. The $f \bar f \to X_1 X_2$ cross section used includes both $\gamma$ and $Z$ exchange, rather than just $\gamma$ exchange given in Eq.~\eqref{3.31}. We have also included all the collision integrals in the numerical calculations, as well as the temperature dependence of $g_\star$. The processes which produce $X_1X_2$ in the final state have one insertion of the dimension-six interaction, so the cross section is proportional to $1/\Lambda^4$. Processes with $X_1X_1$ or $X_2X_2$ in the final state have two insertions of the dimension-six interaction, and are proportional to $1/\Lambda^8$.

The left-hand plot has two light vectors with $m_2=2 m_1$. In this case the freeze-in is dominated by processes that produce $X_1 X_2$ from the scattering of two SM particles.  We evaluated the collision integrals and the relic abundance numerically. For reheat temperatures much larger than $m_1$ and $m_2$, the analytic result in \eqref{3.33} is a very good approximation and the shown curves follow the relation $\Lambda \propto m_1^{1/4}$. For reheat temperatures of the order of $m_1$ or even smaller, the analytic result no longer holds and we observe a qualitatively different behavior of the curves. Note that for very small $T_\text{rh} \lesssim \Lambda_\text{QCD}$ one should not evaluate the collision terms based on quarks, but rather work with hadrons. However, we do not expect the qualitative behavior to change in this region of parameter space.

Certain regions of the plot are excluded. The $X_2$ lifetime should be greater than $10^{26}$~s to avoid distortions of the CMB, or less than $1$~s to avoid affecting the elemental abundances predicted by BBN. We also show the disallowed region from the invisible decay width of the $Z$ and from the limit on the reheat temperature $T_{\text{rh}} < 4.7$~MeV from the CMB and BBN~\cite{deSalas:2015glj}.

The right-hand plot has one heavy vector with a mass $m_2=10$~TeV and one light vector. In this case the dark matter freeze-in is largely determined by the process $\gamma \gamma \to X_1 X_1$ for low reheat temperatures.  Other processes such as $ZZ \to X_1 X_1$ and $\gamma Z \to X_1 X_1$ are relevant if $T_\text{rh}$ is sufficiently large, and are included in the numerics. The $\bar f f \to X_1 X_2$ processes become relevant in the upper part of the plot where $T_\text{rh}$ is only slightly smaller than $m_2$. As for the left-hand plot, we evaluate the collision integrals and the relic abundance numerically. For most  of the plot, the analytical expression \eqref{3.29} is a very good approximation and the curves follow $\Lambda \propto m_1^{1/8}$.

For completeness, we give the $WW \to X_1X_2$ and $Z h \to X_1 X_2$ cross sections in the limit $m_1,m_2 \to 0$, where the expressions simplify considerably:
\begin{align}
\sigma(W^+ W^- \to X_1 X_2) = \frac{e^2 c_W^2 (C_{BX^2}^2+C_{\widetilde BX^2}^2)M_Z^4 s^{1/2} \sqrt{s-4 M_W^2}}{3456  \pi (s - M_Z^2)^2 M_W^4}     (12 M_W^4 + 20 M_W^2 s + s^2)\,,
\end{align}
\begin{align}
\sigma(h Z \to X_1 X_2) &= \frac{ e^2  (C_{BX^2}^2+ C_{\widetilde BX^2}^2) s^2}{1152 \pi  c_W^2 (s-M_Z^2)^2\sqrt{s^2-2s (M_Z^2+M_H^2)+(M_Z^2-M_H^2)^2}} \times \nonumber \\*
& \biggl[ s^2 + 2 s (5 M_Z^2-M_H^2) + (M_H^2-M_Z^2)^2 \biggr]\,.
\end{align}
In the numerics, we have used the complete expressions for the cross sections.

\subsection{Dark Matter Signatures}

Dark matter that is frozen-in is characterized by very feeble interactions with the SM. Models of dark matter freeze-in are therefore generically hard to probe. One expects tiny scattering cross sections of the local dark matter on SM targets, tiny annihilation rates of galactic dark matter to SM particles, and tiny rates of dark matter production at particle colliders. In the following we briefly discuss the scattering of the dark matter on SM fermions in the model presented above.

\begin{figure}[tb]
\centering
\includegraphics[width=6cm]{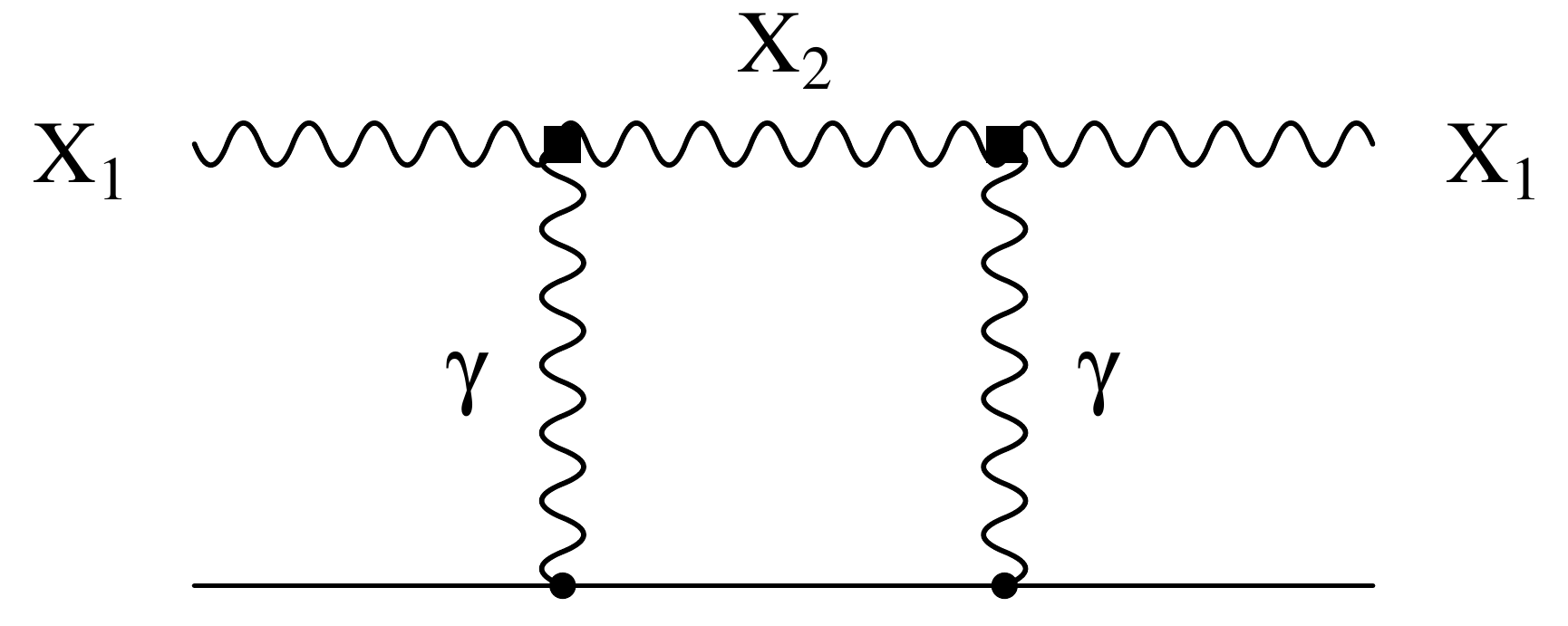}

\caption{Graph contributing to the elastic scattering $X_1+f \to X_1 + f$ off fermions such as electrons or nuclei in the target. The boxes are insertions of the dimension-six interaction.}
\label{fig:elastic}
\end{figure}

Elastic scattering of the dark matter particle $X_1$ on nuclei or electrons can be mediated by a loop of $X_2$ and photons (or $Z$ bosons) that involves two of the dimension 6 interactions, such as the graph shown in Fig.~\ref{fig:elastic}. The corresponding cross sections $\sigma(X_1 N \to X_1 N)$ and $\sigma(X_1 e \to X_1 e)$ are suppressed by a loop factor and eight powers of the new physics scale $\Lambda$. These direct detection cross sections are therefore exceedingly small in the regions of parameter space that give the right dark matter abundance.
%%%%%%%%%%%%%%%%%%%%%%%%%%%%%%%%%
%%%%%%%%%%%%%%%%%%%%%%%%%%%%%%%%%%

If both dark vectors $X_1$ and $X_2$ are almost degenerate, $m_1 \simeq m_2$, one can have inelastic scattering at tree level as shown on the right hand side of Figure~\ref{fig:diagrams}. The corresponding cross section scales as $1/\Lambda^4$. As long as the inelastic processes $X_1 N \to X_2 N$ and $X_1 e \to X_2 e$ are kinematically allowed, they are parametrically much larger than the elastic scattering processes. We first consider the case where $X_2$ is short-lived, so that the dark matter is $X_1$.
As the local dark matter has speeds of the order of $v \sim 10^{-3}$, the mass splitting between $X_2$ and $X_1$ should be at most $\Delta = m_2 - m_1 \sim 10^{-6} m_1$ for inelastic scattering to be kinematically allowed. For inelastic dark matter scattering on a fermion $f$ we find the cross section near threshold
\begin{align}
& \sigma(f + X_1 \to f + X_2) =w \frac{ e^2 Q_f^2 c_W^2 m_1^{1/2} m_2^{5/2} }{24 \pi \sqrt 2 (m_2^2-m_1^2)^{3/2} (m_f+m_2)^3 \sqrt{(2m_f+m_2)^2-m_1^2}}
\biggl\{ \nn
& C_{BX^2}^2 \left(m_2^2-m_1^2 \right)\left[(2m_f+m_2)^2-m_1^2 \right] \left(m_1^2+m_2^2+4m_f^2\right) \nn
&+  C_{\widetilde BX^2}^2  \biggl[ (m_1^2+m_2^2)(m_2^2-m_1^2)^2  +4 m_f m_2 (m_2^2-m_1^2)^2
+ 8 m_f^2 \left( m_2^4 +m_1^4 \right) \nn
& + 16 m_f^3 m_2 (m_1^2+m_2^2) + 16 m_f^4 m_2^2\biggr]  \biggr\}  + \ldots
\end{align}
Here
\begin{align}
s &= (m_f+m_2)^2 + 2 m_f m_1 w^2, & w &\ll 1
\end{align}
is the expansion of $s$ near threshold in terms of  a dimensionless parameter $w$. The relative boost $\gamma$ between $X_1$ and $f$ is
\begin{align}
\gamma &= \frac{p_1 \cdot k_1}{m_1 m_f} =  \frac{m_2^2 -m_1^2 + 2m_f m_2 + 2 m_f m_1 w^2}{2m_1 m_f}
=\frac{m_2^2 -m_1^2 + 2m_f m_2 }{2m_1 m_f} +w^2\,
\end{align}
so $w^2$ is the deviation of $\gamma$ from the minimum value required for $X_2$ production to be kinematically allowed. Since incident dark matter particles are non-relativistic, we can further simplify the cross section by expanding in the mass difference $\Delta=m_2-m_1$, giving
\begin{align}
\sigma(f + X_1 \to f + X_2) &= \frac{ C_{\widetilde BX^2}^2 e^2 Q_f^2 c_W^2 m_f^{3/2} m_1^{7/2} w }{12 \pi \Delta^{3/2} (m_f+m_1)^{3/2}}
+ \ldots
\end{align}
The $t$-channel infrared singularity in the total photon exchange cross section is regulated by the finite mass difference $\Delta$.

For a non-relativistic $X_1$ with velocity $v$ incident on a fermion at rest,
\begin{align}
w^2 &\approx \frac{ 4 m_f m_1 v^2 +m_1^2 v^4 - 8 (m_1+m_f) \Delta-4\Delta^2}{8m_f m_1}\,.
\label{3.40}
\end{align}
As an example, consider the case of $X_1$ with mass $m_1=1\,\text{GeV}$ incident with a typical dark matter velocity $v=10^{-3}$ on a nucleon target with   $m_f=1\,\text{GeV}$.  From Eq.~\eqref{3.40}, $w^2 \approx 5 \times 10^{-7} - 2 \Delta/(1\,\text{GeV})$, and $X_2$ production is kinematically allowed only if $\Delta \le 0.25$~keV. This shows that $X_1$ and $X_2$ have to be almost degenerate for the scattering process to be kinematically allowed for incident particles with typical dark matter velocities. For $\Delta \sim 0.1\, \text{keV}$, $\sigma(f + X_1 \to f + X_2) \approx 5 \times 10^{-24}  \times (1\, \text{GeV}/\Lambda)^4$~$\text{cm}^{2}$ leading to a collision rate $\Gamma \approx (1\, \text{GeV}/\Lambda)^4\times  10^{-22}  \ \text{s}^{-1}$ per target nucleon, assuming $X_1$ is all of the local dark matter density. The nuclear recoil energy is of order $\Delta$, and is approximately 0.1~keV. These numbers show that direct detection of $X_1$ is not a realistic possibility --- $X_1$ and $X_2$ have to be nearly degenerate, the nuclear recoil energy is low, and the collision rate is small for $\Lambda$ much above the electroweak scale.

In some regions of parameter space (the upper part of the left-hand plot in Fig.~\ref{fig:Trh_plots}), $X_2$ is long-lived, and the dark matter is equal parts $X_1$ and $X_2$. In this region, one can also have the exothermic reaction $f+ X_2 \to f+X_1$. We expand the cross section near threshold in $w$ where
\begin{align}
s &= (m_2 + m_f)^2 + 2 m_f m_2 w^2, & w &\ll 1\,.
\end{align}
Here the relative boost of the incoming particles is
\begin{align}
\gamma &= \frac{p_2 \cdot k_1}{m_2 m_f} = 1 + w^2
\end{align}
so that $w^2 \approx v^2/2$ for $v\ll 1$, where $X_2$ particles of velocity $v$ are incident on fermions at rest. The expanded cross section is
\begin{align}
& \sigma(f + X_2 \to f + X_1) = \frac{1}{w} \frac{e^2Q_f^2 c_W^2 m_2\sqrt{(m_2+2m_f)^2-m_1^2}}{192 \pi \sqrt{2} m_f^2  \sqrt{m_2^2-m_1^2} (m_f+m_2)^3} \biggl\{ \nn
&C_{ BX^2}^2 \left[(m_2+2m_f)^2-m_1^2\right] \left[m_2^2-m_1^2 \right] \left[m_1^2+ m_2^2+4m_f^2 \right] +
C_{\widetilde BX^2}^2 \Bigl[ 16 m_f^4 m_2^2 + 16 m_f^3 m_2  (m_1^2+m_2^2) \nn
& + 8 m_f^2  (m_1^4+m_2^4) + 4 m_f m_2  (m_2^2-m_1^2)^2 + (m_2^2-m_1^2)^2 (m_1^2+m_2^2)\Bigr]
\biggr\} + \ldots
\end{align}
For the case of $m_1=1\,\text{GeV}$, $m_2=2\,\text{GeV}$ incident on a nucleon target $m_f=1\,\text{GeV}$,
$\sigma(f + X_2 \to f + X_1) \approx 3 \times 10^{-27}  \times (1\, \text{GeV}/\Lambda)^4$~$\text{cm}^{2}$
leading to a collision rate $\Gamma \approx (1\, \text{GeV}/\Lambda)^4\times 2 \times 10^{-26}  \ \text{s}^{-1}$
assuming that equal amounts of $X_1$ and $X_2$ form the local dark matter density. For $\Lambda \sim 10^{12}\,\text{GeV}$, typical values for $\Lambda$ in this region of parameter space, the reaction rate is too small to be detected in laboratory experiments~\cite{Bloch:2020uzh}.

Another possible detection mechanism is via the process $f + \overline{f} \to X_1 + X_2$ followed by the radiative decay $X_2 \to X_1 + \gamma$, if $X_2$ is short-lived. The cross section is of order $\sigma(f + \overline{f} \to X_1 + X_2 ) \approx   10^{10} \,\text{fb} \times (1\, \text{GeV}/\Lambda)^4$ for $e^+ e^-$ collisions at a center-of-mass energy of $10\,\text{GeV}$. For $\Lambda \sim 10^4 \,\text{GeV}$,  typical values for $\Lambda$ in this region of parameter space, the production rate is too small to be constrained by existing analyses. However, it may be visible in future searches at Belle~II~\cite{Essig:2013vha}.

The above conclusions are a general feature of freeze-in scenarios, where the reaction cross sections required to produce the observed dark matter abundance are small, making them very difficult to detect experimentally.

%%%=====================================================================================================================
\section{Conclusions}\label{sec:conclusions}
%%%=====================================================================================================================

We have presented general extensions of the SMEFT and the LEFT by adding spin 0, 1/2 and 1 particles to the two theories, where the additional particles are singlets under the SM gauge group. We have classified all operators up to dimension-six which involve the new particles, including those which violate baryon and/or lepton number. We have also computed the tree-level matching at the electroweak scale between the two dark matter theories.

An interesting example is the case of vector dark matter with a discrete $\mathbb{Z}_2$ symmetry, where the lightest dark matter particle is stable. Our operator analysis shows that the interaction with SM particles in this case is via a dimension-six triple gauge boson interaction. We have made a preliminary analysis of the phenomenology of this model, and showed that there are regions of parameter space where the freeze-in scenario is viable.

% ============================================================================

\section*{Acknowledgements}
\addcontentsline{toc}{section}{\numberline{}Acknowledgements}

We thank Juan Carlos Criado for useful discussions, and Mukul Sholapurkar for comments on the manuscript. J.\ A.\  acknowledges  financial  support  from  the  European  Research  Council  (ERC)  under the European Union's Horizon 2020 research and innovation programme under grant agreement 833280 (FLAY), and from the Swiss National Science Foundation (SNF) under contract 200020-204428. The research of W.\ A.\ is supported by the U.S. Department of Energy grant number DE-SC0010107. The research of E.\ J.\ and A.\ M.\ is supported by the U.S. Department of Energy DOE grant number DE-SC0009919.

% ============================================================================

\clearpage

\appendix

%%%=====================================================================================================================
\section{Purely dark matter operators}\label{app:DMops}
%%%=====================================================================================================================

In this appendix we list all the self-interactions among the three different DM particles listed in Table~\ref{tab:DMfields} up to dimension six. The generation indices of the DM fields $a,b,c,d$ run over $n_\phi$ scalars, $n_\chi$ fermions, and $n_X$ gauge bosons. The results in this appendix also give the possible interactions up to dimension six in a general gauge theory (with the replacement of derivatives by covariant derivatives).  The possible interactions are restricted by the symmetries of the theory.

The operators range from the dimension-one operator $ \dopp{\phi}{} = \phi_a $, containing only one scalar field $\phi$ up to dimension six. We have chosen to retain the linear term in $\phi_a$, rather than shifting the scalar fields to remove the linear term, to avoid complicated expressions for the matching. In many examples, a linear $\phi_a$ term is forbidden by symmetry. We have also chosen to include the DM kinetic energy terms in the list of dimension-four operators. These terms are usually canonically normalized. Here, integrating out particles at the electroweak scale gives a matching contribution to the DM kinetic energy terms for $X_\mu$. Rather than rescale the DM fields to get a canonically normalized kinetic term, which changes all the operator coefficients, we have instead included the matching in Table~\ref{tab:DMdim1}.

We have given the number of operators of a given type in the tables. In most examples, the counting of operators is trivial. For example, the number of $\phi_a \phi_b \phi_c \phi_d$ operators is $n_\phi(n_\phi+1)(n_\phi+2)(n_\phi+3)/4!$, because the operator is symmetric in $a,b,c,d$. In two cases, the counting is non-trivial. The dimension-six operator with two derivatives and four scalar fields has the form $\dopp{\substack{\partial^2\phi^4 \\ abcd}}{}=\partial_\mu \phi^a \partial^\mu \phi_b \phi_c \phi_d$, which is symmetric in $a,b$ and in $c,d$. $\dopp{\substack{\partial^2\phi^4 \\ abcd}}{}$ then transforms as
\begin{align}
\yng(2) \otimes \yng(2) &= \yng(4 ) \oplus \yng(3,1) \oplus  \yng(2,2)
\label{A.4}
\end{align}
under the permutation group. However, total derivatives do not contribute to the action, so that
\begin{align}
\partial_\mu( \partial^\mu \phi_a \phi_b \phi_c \phi_d)
&= \partial^2 \phi_a \phi_b \phi_c \phi_d +  \partial^\mu  \phi_a \partial_\mu \phi_b \phi_c \phi_d +  \partial^\mu  \phi_a \phi_b \partial_\mu \phi_c \phi_d +  \partial^\mu  \phi_a  \phi_b \phi_c \partial_\mu \phi_d\,,
\label{A.1}
\end{align}
can be set to zero. $\partial^2\phi_a$ can be eliminated by a field-redefinition (i.e.\ the equations of motion), so that $ \partial^\mu  \phi_a \partial_\mu \phi_b \phi_c \phi_d +  \partial^\mu  \phi_a \phi_b \partial_\mu \phi_c \phi_d +  \partial^\mu  \phi_a  \phi_b \phi_c \partial_\mu \phi_d \to 0$ in the action.  As a result, the part of $\dopp{\substack{\partial^2\phi^4 \\ abcd}}{}$ symmetric in three indices vanishes, and $\dopp{\substack{\partial^2\phi^4 \\ abcd}}{}$ transforms as
\begin{equation}\label{A.2}
\dopp{\substack{\partial^2\phi^4 \\ abcd}}{} \sim \yng(2,2)
\end{equation}
under the permutation group, and has dimension
\begin{align}
 \frac1{12} n_\phi^2 (n_\phi^2 -1) \,.
 \label{A.3}
\end{align}
For fermions, there is a Fierz identity
\begin{equation}
  (\chi^T_a C \chi_b)(\chi^T_c C \chi_d)+(\chi^T_a C \chi_c)(\chi^T_d C \chi_b)+(\chi^T_a C \chi_d)(\chi^T_b C \chi_c)=0\,,
  \label{A.5}
\end{equation}
so that $\dopp{\chi\chi}{} $ in Table~\ref{tab:DMdim6} transforms as in Eq.~\eqref{A.2} and has dimension given in Eq.~\eqref{A.3} with $n_\phi \to n_\chi$.

In the matching the vacuum expectation value $v_T$ appears, given by \cite{Jenkins:2013zja,Jenkins:2013wua,Alonso:2013hga}
\begin{equation}
  v_T =\left(1+\frac{3}{8\lambda}C_H v^2\right)v\,.
\end{equation}
Furthermore, we use the notation
\begin{equation}
\begin{pmatrix} a \\ b \end{pmatrix} \equiv \frac{a!}{b!(a-b)!}
\end{equation}
for the binomial coefficient. Note that the DM operators in Tables~\ref{tab:DMdim1}--\ref{tab:DMdim6} are present both in DSMEFT and DLEFT, so what is shown in the tables is the \emph{shift} in the coefficients when electroweak scale particles are integrated out. For example, for the $ \dopp{\phi^4}{} $ operator in Table~\ref{tab:DMdim4}, the coefficient below the EW scale is
\begin{align}
C^{\text{DLEFT}}_{\substack{ \phi^4 \\ abcd}} &= C^{\text{DSMEFT}}_{\substack{ \phi^4 \\ abcd}} + \frac{v_T^2}{2}\Wcci{H^2\phi^4}{abcd}{} \,.
\end{align}
Only the second term is given in Table~\ref{tab:DMdim4}.

%%%
% --- START TABLE
%%%
\begin{table}[H]
\renewcommand{\arraystretch}{1.2}
\small
\begin{align*}
\begin{array}[t]{c|c|c|c|c}
\multicolumn{2}{c}{\textbf{DM: dimension 1}}  \\
\toprule
 (\text{d}_\text{SM}, \text{d}_{\text{DM}})& \text{Name}& \text{Operator} &  \text{Number} &  \text{Matching}  \\
\midrule\midrule
(0,1) & \dopp{\phi}{} & \phi_a & n_\phi & \frac{v_T^2}{2}\left(\Wcci{H^2\phi}{a}{} +\frac{v_T^2}{2}\Wcci{H^4\phi}{a}{}\right) \\
\bottomrule
\end{array}
\end{align*}
\caption{Dimension-one purely DM operator present in DSMEFT and DLEFT. The first column gives the dimensions of the SM and DM part of the operator, the fourth column is the number of operators,  and the fifth column is the \emph{additional} matching contribution at the EW scale.}
\label{tab:DMdim1}
\end{table}
%%%
% --- END TABLE
%%%

%%%
% --- START TABLE
%%%
\begin{table}[H]
\renewcommand{\arraystretch}{1.2}
\small
\begin{align*}
\begin{array}[t]{c|c|c|c|c}
\multicolumn{2}{c}{\textbf{DM: dimension 2}}  \\
\toprule
 (\text{d}_\text{SM}, \text{d}_{\text{DM}})& \text{Name}& \text{Operator} &  \text{Number} &  \text{Matching}  \\
\midrule\midrule
(0,2)  & \dopp{\phi^2}{} & \phi_a\phi_b & \binom{n_\phi+1}{2} & \frac{v_T^2}{2}\left(\Wcci{H^2\phi^2}{ab}{} +\frac{v_T^2}{2}\Wcci{H^4\phi^2}{ab}{}\right) \\
\bottomrule
\end{array}
\end{align*}
\caption{Dimension-two purely DM operator present in DSMEFT and DLEFT. The first column gives the dimensions of the SM and DM part of the operator, the fourth column is the number of operators,  and the fifth column is the \emph{additional} matching contribution  at the EW scale. This operator is the $\phi$ mass term.}
\label{tab:DMdim2}
\end{table}
%%%
% --- END TABLE
%%%

%%%
% --- START TABLE
%%%
\begin{table}[H]
\renewcommand{\arraystretch}{1.2}
\small
\begin{align*}
\begin{array}[t]{c|c|c|c|c}
\multicolumn{2}{c}{\textbf{DM: dimension 3}}  \\
\toprule
 (\text{d}_\text{SM}, \text{d}_{\text{DM}})& \text{Name}& \text{Operator} &  \text{Number} &  \text{Matching}  \\
\midrule\midrule
\multirow{2}{*}{(0,3)} &  \dopp{\phi^3}{} &  \phi_a\phi_b\phi_c &  \binom{n_\phi+2}{3} & \frac{v_T^2}{2}\Wcci{H^2\phi^3}{abc}{} \\
& \dopp{\chi}{} &   (\chi^T_a C \chi_b) + \text{h.c.} &  \binom{n_\chi+1}{2}  + \text{h.c.} & \frac{v_T^2}{2}\Wcci{H\chi}{ab}{} \\
\bottomrule
\end{array}
\end{align*}
\caption{Dimension-three purely DM operators present in DSMEFT and DLEFT. The first column gives the dimensions of the SM and DM part of the operator, the fourth column is the number of operators,  and the fifth column is the \emph{additional} matching contribution  at the EW scale. $\dopp{\chi}{}$ is the $\chi$ mass term.}
\label{tab:DMdim3}
\end{table}
%%%
% --- END TABLE
%%%

%%%
% --- START TABLE
%%%
\begin{table}[H]
\renewcommand{\arraystretch}{1.2}
\small
\begin{align*}
\begin{array}[t]{c|c|c|c|c}
\multicolumn{2}{c}{\textbf{DM: dimension 4}}  \\
\toprule
 (\text{d}_\text{SM}, \text{d}_{\text{DM}})& \text{Name}& \text{Operator} &  \text{Number} &  \text{Matching}  \\
\midrule\midrule
\multirow{6}{*}{(0,4)} & \dopp{\phi}{\text{kin}} &  \partial_\mu \phi_a \partial^\mu \phi_b & \binom{n_\phi +1}{2} & 0  \\
& \dopp{\chi}{\text{kin}} &  \overline \chi_a\, i \slashed{\partial} \, \chi_b & \binom{n_\chi +1}{2} & 0  \\
& \dopp{X}{\text{kin}}&  X_{a\,\mu \nu} X_b^{\mu \nu} & \binom{n_X +1}{2}  & \frac{v_T^2}{2}\Wcci{HX}{ab}{} \\
& \dopp{\widetilde X}{} &  \widetilde X_{a\,\mu \nu} X_b^{\mu \nu} & \binom{n_X +1}{2} & \frac{v_T^2}{2}\Wcci{H\widetilde X}{ab}{} \\
& \dopp{\phi^4}{} &  \phi_a\phi_b\phi_c\phi_d & \binom {n_\phi+3}{4}  & \frac{v_T^2}{2}\Wcci{H^2\phi^4}{abcd}{} \\
& \dopp{\chi\phi}{} &  (\chi_a^T C \chi_b) \phi_c + \text{h.c.} & n_\phi   \binom{n_\chi+1}{2}  + \text{h.c.} & \frac{v_T^2}{2}\Wcci{H\chi\phi}{abc}{} \\
\bottomrule
\end{array}
\end{align*}
\caption{
Dimension-four purely DM operators present in DSMEFT and DLEFT. The first column gives the dimensions of the SM and DM part of the operator, the fourth column is the number of operators,  and the fifth column is the \emph{additional} matching contribution  at the EW scale. The first three operators are the DM kinetic energy terms.}
\label{tab:DMdim4}
\end{table}
%%%
% --- END TABLE
%%%

%%%
% --- START TABLE
%%%
\begin{table}[H]
\renewcommand{\arraystretch}{1.2}
\small
\begin{align*}
\begin{array}[t]{c|c|c|c|c}
\multicolumn{2}{c}{\textbf{DM: dimension 5}}  \\
\toprule
 (\text{d}_\text{SM}, \text{d}_{\text{DM}})& \text{Name}& \text{Operator} &  \text{Number} &  \text{Matching}  \\
\midrule\midrule
\multirow{6}{*}{(0,5)} & \dopp{\phi^5}{} &  \phi_a\phi_b\phi_c\phi_d\phi_e & \binom{n_\phi+4}{5} & 0 \\
& \dopp{X\phi}{} &   X_{a\,\mu \nu} X_b^{\mu \nu} \phi_c & n_\phi \binom{n_X +1}{2}   & 0 \\
& \dopp{\widetilde X\phi}{} &  \widetilde X_{a\,\mu \nu} X_b^{\mu \nu} \phi_c &  n_\phi  \binom{n_X +1}{2}  & 0 \\
& \dopp{\chi \phi^2}{} &    (\chi_a^T C \chi_b) \phi_c\phi_d + \text{h.c.} &  \binom{n_\phi+1}{2}    \binom{n_\chi+1}{2}  + \text{h.c.} & 0 \\
& \dopp{\chi X}{} &   (\chi_a^T C \sigma^{\mu \nu} \chi_b) X_{c\,\mu \nu}  + \text{h.c.} & \binom{n_\chi}{2} n_X + \text{h.c.} & 0 \\
\bottomrule
\end{array}
\end{align*}
\caption{Dimension-five purely DM operators present in DSMEFT and DLEFT. The first column gives the dimensions of the SM and DM part of the operator, the fourth column is the number of operators,  and the fifth column is the \emph{additional} matching contribution  at the EW scale.}
\label{tab:DMdim5}
\end{table}
%%%
% --- END TABLE
%%%

%%%
% --- START TABLE
%%%
\begin{table}[H]
\renewcommand{\arraystretch}{1.2}
\small
\begin{align*}
\begin{array}[t]{c|c|c|c|c}
\multicolumn{2}{c}{\textbf{DM: dimension 6}}  \\
\toprule
 (\text{d}_\text{SM}, \text{d}_{\text{DM}})& \text{Name}& \text{Operator} &  \text{Number} &  \text{Matching}  \\
\midrule\midrule
\multirow{11}{*}{(0,6)} & \dopp{X^3}{} &  X_{a\,\mu}{}^{\nu} X_{b\,\nu }{}^{\alpha} X_{c\,\alpha }{}^{\mu} & \binom{n_X}{3}  &  0\\
& \dopp{\widetilde X^3}{} &   \widetilde X_{a\,\mu}{}^{\nu} X_{b\,\nu }{}^{\alpha} X_{c\,\alpha }{}^{\mu}  & \binom{n_X}{3}  &  0\\
& \dopp{X\phi^2}{} &  X_{a\,\mu \nu} X_b^{\mu \nu} \phi_c\phi_d &  \binom{n_\phi+1}{2}  \binom{n_X +1}{2}   & 0 \\
& \dopp{\widetilde X\phi^2}{} &   \widetilde X_{a\,\mu \nu} X_b^{\mu \nu} \phi_c\phi_d &  \binom{n_\phi+1}{2} \binom{n_X+1}{2}   & 0 \\
& \dopp{\phi^6}{} &   \phi_a\phi_b\phi_c\phi_d\phi_e\phi_f & \binom{n_\phi+5}{6} & 0 \\
& \dopp{\partial^2\phi^4}{} &   \partial_\mu \phi_a \partial^\mu \phi_b \phi_c \phi_d & \frac1{12} n_\phi^2 (n_\phi^2 -1)  &  \frac{\overline g_Z^2}{4M_Z^2}
([Z_\phi]_{ac}-[Z_\phi]_{ca})  ([Z_\phi]_{bd}-[Z_\phi]_{db})\\
& & & & + \,a\leftrightarrow b   \\
& \dopp{\chi\phi^3}{} &   (\chi_a^T C \chi_b) \phi_c\phi_d\phi_e + \text{h.c.} &   \binom{n_\phi +2}{3} \binom{n_\chi+1}{2} + \text{h.c.} & 0\\
& \dopp{\chi X\phi}{} &    (\chi^T_a C \sigma^{\mu \nu} \chi_b) X_{c\,\mu \nu} \phi_d  + \text{h.c.} & n_\phi  \binom{n_\chi}{2} n_X + \text{h.c.} & 0 \\
& \dopp{\chi\partial\phi}{} &  (\overline \chi_a \gamma_\mu \chi_b )  (i\phi_c \overleftrightarrow{\partial^\mu} \phi_d) &  \binom{n_\phi}{2} n_\chi^2 & -\frac{\overline g_2^2}{M_Z^2}[Z_\chi]_{ab} [Z_\phi]_{cd} \\
& \dopp{\chi\chi}{} &   (\chi_a^T C \chi_b)(\chi_c^T C \chi_d)+ \text{h.c.} & \frac1{12} n_\chi^2 (n_\chi^2-1) + \text{h.c.} & 0 \\
& \dopp{\chi\overline\chi}{} &  (\chi_a^T C \chi_b) (\overline \chi_c C \overline \chi_d^T)  &  \binom{n_\chi+1}{2}  \binom{n_\chi+1}{2}  & - \frac{\overline g_Z^2}{M_Z^2} [Z_\chi]_{ca} [Z_\chi]_{db} - \frac{\overline g_Z^2}{M_Z^2} [Z_\chi]_{cb} [Z_\chi]_{da} \\
\bottomrule
\end{array}
\end{align*}
\caption{Dimension-six purely DM operators present in DSMEFT and DLEFT. The first column gives the dimensions of the SM and DM part of the operator, the fourth column is the number of operators,  and the fifth column is the \emph{additional} matching contribution  at the EW scale. }
\label{tab:DMdim6}
\end{table}
%%%
% --- END TABLE
%%%

%\clearpage

%%%=====================================================================================================================
\section{DSMEFT operators}\label{app:DSMEFTops}
%%%=====================================================================================================================

This appendix gives the operators up to dimension six in DSMEFT which have both SM and DM fields. The purely DM operators are given in Appendix~\ref{app:DMops}, and the purely SMEFT operators in Ref.~\cite{Grzadkowski:2010es}. Following the notation in \cite{Grzadkowski:2010es}, we denote the generation indices of the SM fields by $p,r,s,t$ and the number of generations by $n_g$. The subset of operators containing SM fields and a right-handed neutrino has already been studied extensively in the literature. Theories with this field content are called $\nu$SMEFT or SMNEFT and have been discussed for instance in Refs.~\cite{delAguila:2008ir,Aparici:2009fh,Bhattacharya:2015vja,Liao:2016qyd,Bischer:2019ttk,Alcaide:2019pnf}. For this subset of operators, most of the one-loop RG running is known. The gauge mixing is presented in \cite{Datta:2020ocb}, mixing among bosonic operators is discussed in \cite{Chala:2020pbn}, the
one-loop RGE of the four-fermion operators is provided in \cite{Han:2020pff} and the Yukawa mixing for four-fermion operators is  computed recently in \cite{Datta:2021akg}.

It is assumed that all dark matter particles have baryon number $B=0$ and lepton number $L=0$ when we give the $\Delta B$ and $\Delta L$ quantum numbers of the operators. Note that if $\chi$ is a right-handed sterile neutrino, $L=1$, and the lepton number assignments of all the operators get shifted by the number of $\chi$ fields.

%%%
% --- START TABLE
%%%
\begin{table}[H]
\renewcommand{\arraystretch}{1.2}
\small
\begin{align*}
\begin{array}[t]{c|c|c|c}
\multicolumn{2}{c}{\textbf{DSMEFT: dimension 3}} & \multicolumn{2}{c}{\boldsymbol{\Delta B = \Delta L = 0\,}} \\
\toprule
 (\text{d}_\text{SM}, \text{d}_{\text{DM}})& \text{Name}& \text{Operator} &  \text{Number}  \\
\midrule\midrule
(2,1) & \Opp{H^2\phi}{} & (H^\dagger H)\phi_a & n_\phi \\
\bottomrule
\end{array}
\end{align*}
\caption{Dimension-three $\Delta B = \Delta L = 0$ operator in DSMEFT. The first column gives the dimensions of the SM and DM part of the operator, and the fourth column is the number of operators. }
\label{tab:DSMEFTdim3}
\end{table}
%%%
% --- END TABLE
%%%

%%%
% --- START TABLE
%%%
\begin{table}[H]
\renewcommand{\arraystretch}{1.2}
\small
\begin{align*}
\begin{array}[t]{c|c|c|c}
\multicolumn{2}{c}{\textbf{DSMEFT: dimension 4}} & \multicolumn{2}{c}{\boldsymbol{\Delta B = \Delta L = 0\,}} \\
\toprule
 (\text{d}_\text{SM}, \text{d}_{\text{DM}})& \text{Name} & \text{Operator} &  \text{Number}  \\
\midrule\midrule
\multirow{3}{*}{(2,2)} & \Opp{BX}{} & B_{\mu \nu} X_a^{\mu \nu} & n_X  \\
& \Opp{\widetilde BX}{} &  \widetilde B_{\mu \nu} X_a^{\mu \nu} & n_X  \\
& \Opp{H^2\phi^2}{} &  (H^\dagger H )\phi_a\phi_b  &  \binom{n_\phi+1}{2} \\
\bottomrule
\end{array}
\end{align*}
\caption{Dimension-four $\Delta B = \Delta L = 0$ operators in DSMEFT. The first column gives the dimensions of the SM and DM part of the operator, and the fourth column is the number of operators.}
\label{tab:DSMEFTdim4}
\end{table}
%%%
% --- END TABLE
%%%

%%%
% --- START TABLE
%%%
\begin{table}[H]
\renewcommand{\arraystretch}{1.2}
\small
\begin{align*}
\begin{array}[t]{c|c|c|c}
\multicolumn{2}{c}{\textbf{DSMEFT: dimension 5}} & \multicolumn{2}{c}{\boldsymbol{\Delta B = \Delta L = 0\,}} \\
\toprule
 (\text{d}_\text{SM}, \text{d}_{\text{DM}})& \text{Name} & \text{Operator} &  \text{Number}  \\
\midrule\midrule
\multirow{5}{*}{(2,3)} & \Opp{H^2\phi^3}{} &  (H^\dagger H) \phi_a\phi_b\phi_c  &  \binom{n_\phi+2}{3} \\
& \Opp{BX\phi}{} &  B_{\mu \nu} X_a^{\mu \nu} \phi_b  & n_\phi n_X  \\
& \Opp{\widetilde BX\phi}{} &    \widetilde B_{\mu \nu} X_a^{\mu \nu} \phi_b & n_\phi  n_X  \\
& \Opp{H\chi}{} &   (H^\dagger H) (\chi_a^TC\chi_b)  +\text{h.c.} &  \binom{n_\chi+1}{2}  + \text{h.c.}  \\
& \Opp{B\chi}{} &  B_{\mu \nu}  (\chi_a^T C\sigma^{\mu\nu}\chi_b)+ \text{h.c.} & \binom{n_\chi}{2} + \text{h.c.} \\
\hline
\multirow{10}{*}{(4,1)} & \Opp{H^4\phi}{} &  (H^\dagger H)^2 \phi_a & n_\phi \\
& \Opp{B\phi}{} & B_{\mu \nu} B^{\mu \nu} \phi_a & n_\phi \\
& \Opp{\widetilde B\phi}{} & \widetilde B_{\mu \nu} B^{\mu \nu} \phi_a & n_\phi  \\
& \Opp{W\phi}{} & W^I_{\mu \nu} W^{I\,\mu \nu} \phi_a & n_\phi  \\
& \Opp{\widetilde W\phi}{} & \widetilde W^I_{\mu \nu} W^{I\,\mu \nu} \phi_a & n_\phi  \\
& \Opp{G\phi}{} & G^A_{\mu \nu} G^{A\,\mu \nu} \phi_a & n_\phi  \\
& \Opp{\widetilde G\phi}{} & \widetilde G^A_{\mu \nu} G^{A\,\mu \nu} \phi_a & n_\phi  \\
& \Opp{e \phi}{} & (\overline l_p e_r H) \phi_a + \text{h.c.} & n_g^2 n_\phi + \text{h.c.}   \\
& \Opp{u \phi}{} &  (\overline q_p u_r \widetilde H) \phi_a + \text{h.c.} & n_g^2 n_\phi + \text{h.c.}   \\
& \Opp{d \phi}{} & (\overline q_p\, d_r H) \phi_a + \text{h.c.} & n_g^2 n_\phi + \text{h.c.}  \\
\bottomrule
\end{array}
\end{align*}
\caption{Dimension-five $\Delta B = \Delta L = 0$ operators in DSMEFT. The first column gives the dimensions of the SM and DM part of the operator, and the fourth column is the number of operators.}
\label{tab:DSMEFTdim5}
\end{table}
%%%
% --- END TABLE
%%%

%%%
% --- START TABLE
%%%
\begin{table}[H]
\renewcommand{\arraystretch}{1.2}
\small
\begin{align*}
\begin{array}[t]{c|c|c|c}
\multicolumn{2}{c}{\textbf{DSMEFT: dimension 6}} & \multicolumn{2}{c}{\boldsymbol{\Delta B = \Delta L = 0\,}} \\
\toprule
 (\text{d}_\text{SM}, \text{d}_{\text{DM}})& \text{Name} & \text{Operator} &  \text{Number}  \\
\midrule\midrule
\multirow{9}{*}{(2,4)}  & \Opp{BX^2}{} & B_{\mu}^{\,\ \nu} X_{a\,\nu}{}^{\alpha} X_{b\,\alpha}{}^{\mu} & \binom{n_X}{2} \\
& \Opp{\widetilde BX^2}{} &  \widetilde B_{\mu}^{\,\ \nu} X_{a\,\nu}{}^{\alpha} X_{b\,\alpha}{}^{\mu} & \binom{n_X}{2}  \\
& \Opp{BX\phi^2}{} & B_{\mu \nu} X_a^{\mu \nu} \phi_b\phi_c  &  \binom{n_\phi+1}{2} n_X \\
& \Opp{\widetilde BX\phi^2}{} & \widetilde B_{\mu \nu} X_a^{\mu \nu} \phi_b\phi_c &  \binom{n_\phi+1}{2} n_X \\
& \Opp{HX}{} & (H^\dagger H)  X_{a\,\mu \nu} X_b^{\mu \nu}  & \binom{n_X +1}{2}  \\
& \Opp{H\widetilde X}{} &  (H^\dagger H)\widetilde X_{a\,\mu \nu} X_b^{\mu \nu} & \binom{n_X +1}{2}  \\
& \Opp{H^2\phi^4}{} & (H^\dagger H)  \phi_a\phi_b\phi_c\phi_d  &  \binom{n_\phi+3}{4} \\
& \Opp{H\chi\phi}{} & (H^\dagger H) (\chi_a^T C\chi_b) \phi_c + \text{h.c.} &  n_\phi  \binom{n_\chi+1}{2}  + \text{h.c.} \\
& \Opp{B\chi\phi}{} &  B_{\mu \nu}  ( \chi_a^T C \sigma^{\mu \nu} \chi_b) \phi_c  + \text{h.c.} &  n_\phi  \binom{n_\chi}{2} + \text{h.c.} \\
\hline
\multirow{13}{*}{(3,3)}  & \Opp{\partial H\partial\phi}{} & \partial^\mu (H^\dagger H) \partial_\mu (\phi_a \phi_b) & \binom{n_\phi+1}{2} \\
& \Opp{DH\partial\phi}{} & (H^\dagger \overleftrightarrow{D}^\mu H)(\phi_a \overleftrightarrow{\partial_\mu}\phi_b) & \binom{n_\phi}{2} \\
& \Opp{DH\chi}{} &  (iH^\dagger \overleftrightarrow{D}^\mu H)  (\overline \chi_a\gamma_\mu  \chi_b) & n_\chi^2 \\
& \Opp{\phi l}{} &  (\overline l_p \gamma_\mu l_r)  (i\phi_a \overleftrightarrow{\partial^\mu} \phi_b) & n_g^2  \binom{n_\phi}{2}  \\
& \Opp{\phi q}{}& (\overline q_p \gamma_\mu q_r) (i\phi_a \overleftrightarrow{\partial^\mu} \phi_b) & n_g^2    \binom{n_\phi}{2}  \\
& \Opp{\phi e}{} & (\overline e_p \gamma_\mu e_r) (i\phi_a \overleftrightarrow{\partial^\mu} \phi_b) & n_g^2   \binom{n_\phi}{2}  \\
& \Opp{\phi u}{} & (\overline u_p \gamma_\mu u_r)  (i\phi_a \overleftrightarrow{\partial^\mu} \phi_b) & n_g^2   \binom{n_\phi}{2}   \\
& \Opp{\phi d}{} & (\overline d_p \gamma_\mu d_r)  (i\phi_a \overleftrightarrow{\partial^\mu} \phi_b) & n_g^2   \binom{n_\phi}{2} \\
& \Opp{l\chi}{} & (\overline l_p \gamma_\mu l_r) (\overline \chi_a\gamma^\mu \chi_b) & n_g^2 n_\chi^2 \\
& \Opp{q\chi}{} & (\overline q_p \gamma_\mu q_r) (\overline \chi_a\gamma^\mu \chi_b)& n_g^2 n_\chi^2  \\
& \Opp{e\chi}{} &  (\overline e_p \gamma_\mu e_r) (\overline \chi_a\gamma^\mu \chi_b) & n_g^2 n_\chi^2 \\
& \Opp{u\chi}{} & (\overline u_p \gamma_\mu u_r) (\overline \chi_a\gamma^\mu \chi_b) & n_g^2 n_\chi^2 \\
& \Opp{d\chi}{} & (\overline d_p \gamma_\mu d_r) (\overline \chi_a\gamma^\mu \chi_b) & n_g^2 n_\chi^2  \\
\bottomrule
\end{array}
\end{align*}
\caption{Dimension-six $\Delta B = \Delta L = 0$ operators in DSMEFT, part 1. The first column gives the dimensions of the SM and DM part of the operator, and the fourth column is the number of operators.}
\label{tab:DSMEFTdim61}
\end{table}
%%%
% --- END TABLE
%%%

%%%
% --- START TABLE
%%%
\begin{table}[H]
\renewcommand{\arraystretch}{1.2}
\small
\begin{align*}
\begin{array}[t]{c|c|c|c}
\multicolumn{2}{c}{\textbf{DSMEFT: dimension 6}} & \multicolumn{2}{c}{\boldsymbol{\Delta B = \Delta L = 0\,}} \\
\toprule
 (\text{d}_\text{SM}, \text{d}_{\text{DM}})& \text{Name} & \text{Operator} &  \text{Number}  \\
\midrule\midrule
\multirow{17}{*}{(4,2)}
& \Opp{H^4\phi^2}{} & (H^\dagger H)^2 \phi_a\phi_b & \binom{n_\phi+1}{2} \\
& \Opp{B\phi^2}{} & B_{\mu \nu} B^{\mu \nu} \phi_a\phi_b &  \binom{n_\phi+1}{2}  \\
& \Opp{\widetilde B\phi^2}{} & \widetilde B_{\mu \nu} B^{\mu \nu} \phi_a\phi_b & \binom{n_\phi+1}{2}  \\
& \Opp{W\phi^2}{} &  W^I_{\mu \nu} W^{I \mu \nu} \phi_a\phi_b &  \binom{n_\phi+1}{2}  \\
& \Opp{\widetilde W\phi^2}{} & \widetilde W^I_{\mu \nu} W^{I\,\mu \nu} \phi_a\phi_b &  \binom{n_\phi+1}{2}  \\
& \Opp{G\phi^2}{} & G^A_{\mu \nu} G^{A\,\mu \nu} \phi_a\phi_b &  \binom{n_\phi+1}{2}   \\
& \Opp{\widetilde G\phi^2}{} &  \widetilde G^A_{\mu \nu} G^{A\,\mu \nu} \phi_a\phi_b &  \binom{n_\phi+1}{2}  \\
& \Opp{H^2BX}{} & (H^\dagger H) B_{\mu \nu} X_a^{\mu \nu} & n_X \\
& \Opp{H^2\widetilde BX}{} & (H^\dagger H) \widetilde B_{\mu \nu} X_a^{\mu \nu} & n_X \\
& \Opp{H^2WX}{} &  (H^\dagger\tau^I H) W^I_{\mu \nu} X_a^{\mu \nu} & n_X  \\
& \Opp{H^2\widetilde WX}{} & (H^\dagger\tau^I H) \widetilde W^I_{\mu \nu} X_a^{\mu \nu}  & n_X \\
& \Opp{e\phi^2}{}  & (\overline l_p e_r H) \phi_a\phi_b + \text{h.c.} & n_g^2   \binom{n_\phi+1}{2}  + \text{h.c.} \\
& \Opp{u\phi^2}{} & (\overline q_p u_r \widetilde H) \phi_a\phi_b + \text{h.c.} & n_g^2   \binom{n_\phi+1}{2}  + \text{h.c.} \\
& \Opp{d\phi^2}{} & (\overline q_p\, d_r H) \phi_a\phi_b + \text{h.c.} & n_g^2 \binom{n_\phi+1}{2}  + \text{h.c.} \\
& \Opp{eX}{} &  (\overline l_p\sigma_{\mu \nu} e_r)H X_{a}^{\mu \nu} + \text{h.c.}  & n_g^2 n_X + \text{h.c.}  \\
& \Opp{uX}{} & (\overline q_p\sigma_{\mu\nu} u_r)\widetilde{H} X_{a}^{\mu \nu}+ \text{h.c.} & n_g^2 n_X + \text{h.c.}  \\
& \Opp{dX}{} &  (\overline q_p\sigma_{\mu\nu} d_r)H X_{a}^{\mu \nu} + \text{h.c.} & n_g^2 n_X + \text{h.c.} \\
\bottomrule
\end{array}
\end{align*}
\caption{Dimension-six $\Delta B = \Delta L = 0$ operators in DSMEFT, part 2. The first column gives the dimensions of the SM and DM part of the operator, and the fourth column is the number of operators.}
\label{tab:DSMEFTdim62}
\end{table}
%%%
% --- END TABLE
%%%

%%%
% --- START TABLE
%%%
\begin{table}[H]
\renewcommand{\arraystretch}{1.2}
\small
\begin{align*}
\begin{array}[t]{c|c|c|c}
\multicolumn{2}{c}{\textbf{DSMEFT: dimension 4}} & \multicolumn{2}{c}{\boldsymbol{\Delta B=0,\ \Delta L =1 + \text{h.c.}\,}} \\
\toprule
 (\text{d}_\text{SM}, \text{d}_{\text{DM}})& \text{Name}& \text{Operator} &  \text{Number}  \\
\midrule\midrule
(5/2,3/2) & \Opp{H\chi l}{} &  \epsilon_{ij} H^i  (\overline \chi_a l_p^j) & n_g n_\chi \\
\bottomrule
\end{array}
\end{align*}
\caption{Dimension-four $\Delta B=0$, $\Delta L =1$ operator in DSMEFT. The first column gives the dimensions of the SM and DM part of the operator, and the fourth column is the number of operators.}
\label{tab:DSMEFTdim4DL1}
\end{table}
%%%
% --- END TABLE
%%%

%%%
% --- START TABLE
%%%
\begin{table}[H]
\renewcommand{\arraystretch}{1.2}
\small
\begin{align*}
\begin{array}[t]{c|c|c|c}
\multicolumn{2}{c}{\textbf{DSMEFT: dimension 5}} & \multicolumn{2}{c}{\boldsymbol{\Delta B=0,\ \Delta L =1 + \text{h.c.}\,}} \\
\toprule
 (\text{d}_\text{SM}, \text{d}_{\text{DM}})& \text{Name}& \text{Operator} &  \text{Number}  \\
\midrule\midrule
(5/2,5/2)  &  \Opp{H\chi l\phi}{} &\epsilon_{ij} H^i  (\overline \chi_a l_p^j)   \phi_b & n_g  n_\phi  n_\chi \\
\bottomrule
\end{array}
\end{align*}
\caption{Dimension-five $\Delta B=0$, $\Delta L =1$ operator in DSMEFT. The first column gives the dimensions of the SM and DM part of the operator, and the fourth column is the number of operators.}
\label{tab:DSMEFTdim5DL1}
\end{table}
%%%
% --- END TABLE
%%%

%%%
% --- START TABLE
%%%
\begin{table}[H]
\renewcommand{\arraystretch}{1.2}
\small
\begin{align*}
\begin{array}[t]{c|c|c|c}
\multicolumn{2}{c}{\textbf{DSMEFT: dimension 6}} & \multicolumn{2}{c}{\boldsymbol{\Delta B=0,\ \Delta L =1 + \text{h.c.}\,}} \\
\toprule
 (\text{d}_\text{SM}, \text{d}_{\text{DM}})& \text{Name}& \text{Operator} &  \text{Number}  \\
\midrule\midrule
\multirow{3}{*}{(5/2,7/2)} & \Opp{H\chi l\phi^2}{} & \epsilon_{ij} H^i   (\overline \chi_a l_p^j) \phi_b\phi_c  &  n_g   \binom{n_\phi+1}{2}  n_\chi  \\
& \Opp{H\chi l\partial \phi}{} &   \epsilon_{ij} H^i (\chi^T_a C \gamma_\mu l_p^j) \partial^\mu \phi_b & n_g n_\phi n_\chi  \\
& \Opp{H\chi lX}{} & \epsilon_{ij} H^i   ( \overline \chi_a\sigma^{\mu\nu} l_p^j) X_{b\,\mu \nu} & n_g n_\chi n_X  \\
\hline
\multirow{10}{*}{(9/2,3/2)} & \Opp{H^3\chi l}{} &  \epsilon_{ij} H^i (H^\dagger H) (\overline \chi_a l_p^j) & n_g n_\chi  \\
& \Opp{H\chi e}{} & (i\widetilde H^\dagger D^\mu H)  (\overline \chi_a \gamma_\mu  e_p) & n_g n_\chi \\
& \Opp{HB\chi l}{} & \epsilon_{ij}H^j  B_{\mu \nu} ( \overline \chi_a \sigma^{\mu\nu} l_p^i) & n_g n_\chi \\
& \Opp{HW\chi l}{} & \epsilon_{ik}(\tau^I)^k{}_j H^jW^I_{\mu \nu}( \overline \chi_a \sigma^{\mu\nu} l_p^i) & n_g n_\chi\\
& \Opp{le\chi}{(1)} & \epsilon_{ij}  (l_p^{i\,T} C l_r^j)  (\overline e_s C \overline \chi^T_a)  & n_g \binom{n_g}{2}   n_\chi \\
& \Opp{le\chi}{(3)} &  \epsilon_{ij} (l_p^{i\,T} C \sigma^{\mu \nu}  l_r^j)  (\overline e_s \sigma_{\mu \nu} C \overline \chi^T_a)  & n_g \binom{n_g+1}{2} n_\chi \\
& \Opp{dq\chi l}{(1)} & \epsilon_{ij}(\overline d_p q_r^i)  (\overline \chi_a l_s^j)  & n_g^3 n_\chi \\
& \Opp{dq\chi l}{(3)} & \epsilon_{ij}(\overline d_p \sigma_{\mu \nu} q_r^i)  (\overline \chi_a \sigma^{\mu \nu}  l_s^j)  &n_g^3 n_\chi \\
& \Opp{du\chi e}{}& (\overline d_p \gamma_\mu u_r) (\overline \chi_a \gamma^\mu e_s) & n_g^3 n_\chi \\
& \Opp{qu\chi l}{} & (\overline q_p^i  u_r) (\overline \chi_a  l_s^i) & n_g^3 n_\chi \\
\bottomrule
\end{array}
\end{align*}
\caption{Dimension-six $\Delta B=0$, $\Delta L =1$ operators in DSMEFT. The first column gives the dimensions of the SM and DM part of the operator, and the fourth column is the number of operators.}
\label{tab:DSMEFTdim6DL1}
\end{table}
%%%
% --- END TABLE
%%%

%%%
% --- START TABLE
%%%
\begin{table}[H]
\renewcommand{\arraystretch}{1.2}
\small
\begin{align*}
\begin{array}[t]{c|c|c|c}
\multicolumn{2}{c}{\textbf{DSMEFT: dimension 6}} & \multicolumn{2}{c}{\boldsymbol{\Delta B=0,\ \Delta L =2 + \text{h.c.}\,}} \\
\toprule
 (\text{d}_\text{SM}, \text{d}_{\text{DM}})& \text{Name}& \text{Operator} &  \text{Number}  \\
\midrule\midrule
(5,1) & \Opp{\nu\nu\phi}{} & \epsilon_{ij} \epsilon_{kl} H^j H^l (l_p^{i\,T} C l_r^{k })   \phi_a & \binom{n_g+1}{2}  n_\phi \\
\bottomrule
\end{array}
\end{align*}
\caption{Dimension-six $\Delta B=0$, $\Delta L =2$ operator in DSMEFT. The first column gives the dimensions of the SM and DM part of the operator, and the fourth column is the number of operators.}
\label{tab:DSMEFTDL2}
\end{table}
%%%
% --- END TABLE
%%%

%%%
% --- START TABLE
%%%
\begin{table}[H]
\renewcommand{\arraystretch}{1.2}
\small
\begin{align*}
\begin{array}[t]{c|c|c|c}
\multicolumn{2}{c}{\textbf{DSMEFT: dimension 6}} & \multicolumn{2}{c}{\boldsymbol{\Delta L=0,\ \Delta B=1 + \text{h.c.}\,}} \\
\toprule
 (\text{d}_\text{SM}, \text{d}_{\text{DM}})& \text{Name}& \text{Operator} &  \text{Number}  \\
\midrule\midrule
\multirow{3}{*}{(9/2,3/2)} & \Opp{ddu}{(1)} & \epsilon^{\alpha\beta\gamma}(d_p^{\alpha\, T} C d_r^\beta) (\chi^T_a C u_s^{\gamma}) & n_g \binom{n_g}{2}  n_\chi \\
& \Opp{ddu}{(3)} & \epsilon^{\alpha\beta\gamma}(d_p^{\alpha\, T} C \sigma_{\mu \nu} d_r^\beta) (\chi^T_a C \sigma^{\mu \nu} u_s^{\gamma}) & n_g \binom{n_g+1}{2}  n_\chi \\
& \Opp{qqd}{} &  \epsilon^{\alpha\beta\gamma}\epsilon_{ij}(q_p^{i\alpha\,T}C q_r^{j\beta}) (\chi^T_a C d_s^\gamma) & n_g \binom{n_g+1}{2} n_\chi \\
\bottomrule
\end{array}
\end{align*}
\caption{Dimension-six $\Delta L=0$, $\Delta B=1$ operators in DSMEFT. The first column gives the dimensions of the SM and DM part of the operator, and the fourth column is the number of operators.}
\label{tab:DSMEFTDB1}
\end{table}
%%%
% --- END TABLE
%%%

%\clearpage

%=====================================================================================================================
 \section{DLEFT operators}\label{app:DLEFTops}
%=====================================================================================================================

This appendix gives the DLEFT operators up to dimension six that include both DM and LEFT fields. The purely DM operators are given in Appendix~\ref{app:DMops} and the purely LEFT operators in \cite{Jenkins:2017jig,Jenkins:2017dyc}. We also give the tree-level matching contributions to these operators from DSMEFT, when heavy SM degrees of freedom are integrated out at the EW scale. The matching conditions depend on the $\mathcal{Z}$ and $\mathcal{W}$ couplings
\begin{align}
  [Z_\phi]_{ab}&=\frac12 v_T^2\,C_{\substack{DH\partial\phi \\ ab}}\,,&
  [Z_\chi]_{ab}&=-\frac12 v_T^2\,C_{\substack{DH\chi \\ ab}}\,, &
  [W_\chi]_{ap}&=\frac12 v_T^2\,C_{\substack{H\chi e\\ ap}}\,, &
\end{align}
and
\begin{align}
  A_{BW} & = v_T^2  \overline s\,\Wcc{H^2BX}{} + v_T^2 \overline c\,\Wcc{H^2WX}{} + 2 \overline s \Wcc{BX}{} \,, \\
  A_{\widetilde B W} & = v_T^2 \overline s\,\Wcc{H^2\widetilde BX}{} + v_T^2 \overline c\,\Wcc{H^2\widetilde WX}{} + 2 \overline s \Wcc{\widetilde BX}{} \,,
\end{align}
The neutral and charged currents are
\begin{align}
J^\mu &= A_{BW} \partial_\nu X^{\nu \mu} + A_{\widetilde BW} \partial_\nu \widetilde X^{\nu \mu} ,\\
  j_Z^\mu &= [Z_\phi]_{ab}(i\phi_a\overleftrightarrow{\partial^\mu} \phi_b)+[Z_\chi]_{ab}(\overline \chi_a \gamma^\mu \chi_b)+[Z_{\nu_L}]_{pr}(\overline \nu_{Lp}\gamma^\mu\nu_{Ls})+[Z_{e_L}]_{pr}(\overline e_{Lp}\gamma^\mu e_{Lr})\nonumber \\
  &+[Z_{e_R}]_{pr}(\overline e_{Rp}\gamma^\mu e_{Rr})+[Z_{u_L}]_{pr}(\overline u_{Lp}\gamma^\mu u_{Lr})+[Z_{u_R}]_{pr}(\overline u_{Rp}\gamma^\mu u_{Rr})+[Z_{d_L}]_{pr}(\overline d_{Lp}\gamma^\mu d_{Lr})\nn
  &+[Z_{d_R}]_{pr}(\overline d_{Rp}\gamma^\mu d_{Rr})\,,\\
 j_W^\mu &= [W_\chi]_{ap}(\overline \chi_a \gamma^\mu e_p)+[W_l]_{pr}(\overline \nu_{Lp} \gamma^\mu e_{Lr})+[W_q]_{pr}(\overline u_{Lp} \gamma^\mu d_{Lr})+[W_R]_{pr}(\overline u_{Rp} \gamma^\mu d_{Rr})\,,
\end{align}
which enter the interaction Lagrangian
\begin{equation}
  \mathcal{L} = -\mathcal{Z}_\mu \left(\overline g_Z  j^\mu_Z+J^\mu \right)-\frac{\overline g_2}{\sqrt{2}}\left(\mathcal{W}^+_\mu j^\mu_W+\text{h.c.}\right) + \ldots \,.
\end{equation}
The relation of the mass eigenstate fields $\mathcal{Z}$ and $\mathcal{W}$ to the electroweak fields $W^3$ and $B$ as well as the couplings $[Z_{f_i}]_{pr}$ for SM fermions
are defined in \cite{Jenkins:2017jig}. The couplings $\overline g_Z$ and $\overline g_W$ are defined in Ref.~\cite{Alonso:2013hga}.
Integrating out the $\mathcal{Z}$ and $\mathcal{W}$ fields give the low-energy interaction
\begin{align}
\mathcal{L} &= -\frac{\overline g_2^2}{2 M_W^2} j_{W\mu} j^\mu_W-\frac{\overline g_Z^2}{2 M_Z^2} j_{Z\mu} j^\mu_Z -\frac{\overline g_Z }{M_Z^2} j_{Z\mu} J^\mu  -\frac{ 1}{2 M_Z^2} J^\mu J_\mu\,.
\label{c7}
\end{align}
The $\widetilde X$ term in $J^\mu$ can be dropped using  the Bianchi identity
\begin{align}
\partial_\nu \widetilde X^{\mu \nu} &=0\,.
\end{align}
The equation of motion $\partial_\mu X^{\mu \nu} = j_X^\nu$ can be used to rewrite Eq.~\eqref{c7} as
\begin{align}
\mathcal{L} &= -\frac{\overline g_2^2}{2 M_W^2} j_{W\mu} j^\mu_W-\frac{\overline g_Z^2}{2 M_Z^2} j_{Z\mu} j^\mu_Z + \frac{\overline g_Z A_{BW} }{M_Z^2} j_{Z\mu} j^\mu_X  -\frac{A^2_{BW}}{2 M_Z^2} j^\mu_X j_{X\mu} \,.
\label{c9}
\end{align}
The first two terms give the matching contributions listed in the tables. The last two terms depend on the dark matter currents $j^\mu_X$, which are given in terms of the DM gauge generators, which we have not specified. Their contribution must be added to the matching.

There are also matching contributions from Higgs exchange. However, as discussed in \cite{Jenkins:2017jig}, there are additional suppressions to Higgs couplings because, by assumption, the particles in DLEFT all have masses much smaller than the electroweak scale. The Yukawa couplings of the light SM fermions are $m_{\text{light}}/v$, and formally one higher order in the DLEFT power counting. There are similar suppressions of $\phi_a-H$ and $\chi_a-H$ interactions, so that $\phi_a$ and $\chi_a$ remain light after electroweak symmetry breaking. These are:
\begin{enumerate}
\item $\phi (H^\dagger H)$ has a $(m_{\text{light}}/v)^3$ suppression.
\item $\phi^2 (H^\dagger H)$ has a $(m_{\text{light}}/v)^2$ suppression.
\item $\phi (H^\dagger H)^2$ has a $(m_{\text{light}}/v)^3$ suppression.
\item $\phi^2 (H^\dagger H)^2 $ has a $(m_{\text{light}}/v)^2$ suppression.
\item $\phi^3 (H^\dagger H) $ has a $m_{\text{light}}/v$ suppression.
\item $\chi^2 (H^\dagger H) $ has a $m_{\text{light}}/v$ suppression.
\item $H \overline \chi l $ has a $m_{\text{light}}/v$ suppression.
\item $H (H^\dagger H)  \overline \chi l $ has a $m_{\text{light}}/v$ suppression.
\end{enumerate}
As a result, Higgs exchange does not contribute at dimension-six, and can be neglected, as was the case in LEFT \cite{Jenkins:2017jig}.

%%%
% --- START TABLE
%%%
\begin{table}[H]
\renewcommand{\arraystretch}{1.5}
\small
\begin{align*}
\begin{array}[t]{c|c|c|c|c}
\multicolumn{2}{c}{\textbf{DLEFT: dimension 4}} & \multicolumn{3}{c}{\boldsymbol{\Delta B = \Delta L = 0\,}} \\
\toprule
 (\text{d}_\text{SM}, \text{d}_{\text{DM}})& \text{Name} & \text{Operator} &  \text{Number} &  \text{Matching}  \\
\midrule\midrule
\multirow{2}{*}{(2,2)}  & \opp{FX}{} & F_{\mu \nu} X_a^{\mu \nu} & n_X & \overline c\, \Wcci{BX}{a}{}+\frac{v_T^2}{2}\left(\overline c\, \Wcci{H^2BX}{a}{}-\overline s\, \Wcci{H^2WX}{a}{}\right)\\
& \opp{\widetilde FX}{} &  \widetilde F_{\mu \nu} X_a^{\mu \nu} & n_X & \overline c\, \Wcci{\widetilde BX}{a}{} +\frac{v_T^2}{2}\left(\overline c\, \Wcci{H^2\widetilde BX}{a}{}-\overline s\, \Wcci{H^2\widetilde WX}{a}{}\right) \\
\hline
\multirow{3}{*}{(3,1)} & \opp{e\phi}{} & (\overline e_{L\,p} e_{R\,r}) \phi_a + \text{h.c.} & n_g^2 n_\phi + \text{h.c.}  & \frac{v_T}{\sqrt{2}}\Wcci{e\phi}{pra}{} \\
& \opp{u\phi}{} & (\overline u_{L\,p} u_{R\,r}) \phi_a + \text{h.c.} & n_g^2 n_\phi + \text{h.c.}  & \frac{v_T}{\sqrt{2}}\Wcci{u\phi}{pra}{} \\
& \opp{d\phi }{} & (\overline d_{L\,p} d_{R\,r}) \phi_a + \text{h.c.} & n_g^2 n_\phi + \text{h.c.} & \frac{v_T}{\sqrt{2}}\Wcci{d\phi}{pra}{} \\
\bottomrule
\end{array}
\end{align*}
\caption{Dimension-four $\Delta B = \Delta L = 0$ operators in DLEFT.  Note that $ \opp{e\phi}{} $ is not the same as $\Opp{e \phi}{} $, etc. The first column gives the dimensions of the SM and DM part of the operator, the fourth column is the number of operators,  and the fifth column is the additional matching contribution  at the EW scale.
}
\label{tab:DLEFTdim4}
\end{table}
%%%
% --- END TABLE
%%%

%%%
% --- START TABLE
%%%
\begin{table}[H]
\renewcommand{\arraystretch}{1.5}
\small
\begin{align*}
\begin{array}[t]{c|c|c|c|c}
\multicolumn{2}{c}{\textbf{DLEFT: dimension 5}} & \multicolumn{3}{c}{\boldsymbol{\Delta B = \Delta L = 0\,}} \\
\toprule
 (\text{d}_\text{SM}, \text{d}_{\text{DM}})& \text{Name} & \text{Operator} &  \text{Number} &  \text{Matching}  \\
\midrule\midrule
\multirow{3}{*}{(2,3)} & \opp{FX\phi}{} &  F_{\mu \nu} X_a^{\mu \nu} \phi_b  & n_\phi  n_X & \overline c\, \Wcci{BX\phi}{ab}{} \\
& \opp{\widetilde FX\phi}{} & \widetilde F_{\mu \nu} X_a^{\mu \nu} \phi_b & n_\phi n_X  & \overline c\, \Wcci{\widetilde BX\phi}{ab}{}  \\
& \opp{F\chi}{} & F_{\mu \nu}  (\chi_a^T C\sigma^{\mu \nu}\chi_b)  + \text{h.c.} & \binom{n_\chi}{2}  + \text{h.c.} & \overline c\, \Wcci{B\chi}{ab}{} \\
\hline
\multirow{6}{*}{(3,2)} & \opp{e\phi^2 }{} & (\overline e_{L\,p} e_{R\,r}) \phi_a\phi_b+ \text{h.c.} & n_g^2  \binom{n_\phi+1}{2} + \text{h.c.}  & \frac{v_T}{\sqrt{2}}\Wcci{e\phi^2}{prab}{} \\
& \opp{u\phi^2}{} & (\overline u_{L\,p} u_{R\,r}) \phi_a\phi_b + \text{h.c.} & n_g^2  \binom{n_\phi+1}{2} + \text{h.c.}  & \frac{v_T}{\sqrt{2}}\Wcci{u\phi^2}{prab}{} \\
& \opp{d\phi^2}{} & (\overline d_{L\,p} d_{R\,r}) \phi_a\phi_b + \text{h.c.}  & n_g^2   \binom{n_\phi+1}{2}  + \text{h.c.}  & \frac{v_T}{\sqrt{2}}\Wcci{d\phi^2}{prab}{} \\
& \opp{eX}{} & (\overline e_{L\,p} \sigma_{\mu \nu}e_{R\,r}) X_{a}^{\mu \nu} + \text{h.c.} &  n_g^2 n_X + \text{h.c.} & \frac{v_T}{\sqrt{2}}\Wcci{eX}{pra}{} \\
& \opp{uX}{} & (\overline u_{L\,p} \sigma_{\mu \nu}u_{R\,r}) X_{a}^{\mu \nu} + \text{h.c.} &  n_g^2 n_X + \text{h.c.} & \frac{v_T}{\sqrt{2}}\Wcci{uX}{pra}{} \\
& \opp{dX}{} & (\overline d_{L\,p} \sigma_{\mu \nu}d_{R\,r}) X_{a}^{\mu \nu} + \text{h.c.} & n_g^2 n_X + \text{h.c.} & \frac{v_T}{\sqrt{2}}\Wcci{dX}{pra}{} \\
\hline
\multirow{4}{*}{(4,1)} & \opp{F\phi}{} & F_{\mu \nu} F^{\mu \nu} \phi_a & n_\phi & \overline c^2\, \Wcci{B\phi}{a}{}+\overline s^2\, \Wcci{W\phi}{a}{} \\
& \opp{\widetilde F\phi}{} & \widetilde F_{\mu \nu} F^{\mu \nu} \phi_a  & n_\phi & \overline c^2\, \Wcci{\widetilde B\phi}{a}{} +\overline s^2\, \Wcci{\widetilde W\phi}{a}{} \\
& \opp{G\phi}{} & G^A_{\mu \nu} G^{A\,\mu \nu} \phi_a  & n_\phi & \Wcci{G\phi}{a}{} \\
& \opp{\widetilde G\phi}{} & \widetilde G^A_{\mu \nu} G^{A\,\mu \nu} \phi_a & n_\phi & \Wcci{\widetilde G\phi}{a}{} \\
\bottomrule
\end{array}
\end{align*}
\caption{Dimension-five $\Delta B = \Delta L = 0$ operators in DLEFT.  Note that $ \opp{e\phi^2}{} $ and $\opp{eX}{}$ are not the same as $\Opp{e \phi^2}{} $ and $\Opp{eX}{}$, etc.  The first column gives the dimensions of the SM and DM part of the operator, the fourth column is the number of operators,  and the fifth column is the additional matching contribution  at the EW scale.}
\label{tab:DLEFTdim5}
\end{table}
%%%
% --- END TABLE
%%%

%%%
% --- START TABLE
%%%
\begin{table}[H]
\renewcommand{\arraystretch}{1.5}
\small
\begin{align*}
\begin{array}[t]{c|c|c|c|c}
\multicolumn{2}{c}{\textbf{DLEFT: dimension 6}} & \multicolumn{3}{c}{\boldsymbol{\Delta B = \Delta L = 0\,}} \\
\toprule
 (\text{d}_\text{SM}, \text{d}_{\text{DM}})& \text{Name} & \text{Operator} &  \text{Number} &  \text{Matching}  \\
\midrule\midrule
\multirow{5}{*}{(2,4)}  & \opp{FX^2}{} & F_{\mu }^{\,\,\nu} X_{a\,\nu }{}^{\alpha} X_{b\,\alpha }{}^{\mu} & \binom{n_X}{2} & \overline c\, \Wcci{BX^2}{ab}{} \\
& \opp{\widetilde FX^2}{} & \widetilde F_{\mu}^{\,\,\,\nu} X_{a\,\nu }{}^{\alpha} X_{b\,\alpha }{}^{\mu}  &  \binom{n_X}{2} & \overline c\, \Wcci{\widetilde BX^2}{ab}{} \\
& \opp{FX\phi^2}{} & F_{\mu \nu} X_a^{\mu \nu} \phi_b\phi_c &  \binom{n_\phi+1}{2} n_X  & \overline c\, \Wcci{BX\phi^2}{abc}{} \\
& \opp{\widetilde FX\phi^2}{} &  \widetilde F_{\mu \nu} X_a^{\mu \nu} \phi_b\phi_c &  \binom{n_\phi+1}{2}  n_X & \overline c\, \Wcci{\widetilde BX\phi^2}{abc}{} \\
& \opp{F\chi\phi}{} &  F_{\mu \nu} (\chi_a^T C\sigma^{\mu \nu}\chi_b)\phi_c + \text{h.c.} & n_\phi   \binom{n_\chi}{2}   + \text{h.c.} & \overline c\, \Wcci{B\chi\phi}{abc}{} \\
\bottomrule
\end{array}
\end{align*}
\caption{Dimension-six $\Delta B = \Delta L = 0$ operators in DLEFT, part 1. The first column gives the dimensions of the SM and DM part of the operator, the fourth column is the number of operators,  and the fifth column is the additional matching contribution  at the EW scale.}
\label{tab:DLEFTdim61}
\end{table}
%%%
% --- END TABLE
%%%

%%%
% --- START TABLE
%%%
\begin{table}[H]
\renewcommand{\arraystretch}{1.5}
\small
\begin{align*}
\begin{array}[t]{c|c|c|c|c}
\multicolumn{2}{c}{\textbf{DLEFT: dimension 6}} & \multicolumn{3}{c}{\boldsymbol{\Delta B = \Delta L = 0\,}} \\
\toprule
 (\text{d}_\text{SM}, \text{d}_{\text{DM}})& \text{Name} & \text{Operator} &  \text{Number} &  \text{Matching}  \\
\midrule\midrule
\multirow{13}{*}{(3,3)} & \opp{e\phi^3}{} & (\overline e_{L\,p} e_{R\,r}) \phi_a\phi_b\phi_c + \text{h.c.} & n_g^2 \binom{n_\phi+2}{3}  +\text{h.c.} & 0 \\
& \opp{u\phi^3 }{} &  (\overline u_{L\,p} u_{R\,r}) \phi_a\phi_b\phi_c + \text{h.c.} & n_g^2 \binom{n_\phi+2}{3}  +\text{h.c.} & 0 \\
& \opp{d\phi^3 }{} &  (\overline d_{L\,p} d_{R\,r}) \phi_a\phi_b\phi_c + \text{h.c.} & n_g^2 \binom{n_\phi+2}{3}  +\text{h.c.} & 0 \\
& \opp{\phi \nu}{L}   & (\overline \nu_{L\,p} \gamma_\mu \nu_{L\,r})  (i\phi_a \overleftrightarrow{\partial^\mu} \phi_b) &  n_g^2  \binom{n_\phi}{2}  & \Wcci{\phi l}{prab}{} -\frac{\overline g_Z^2}{M_Z^2}[Z_{\nu}]_{pr}[Z_\phi]_{ab}\\
& \opp{\phi e}{L} & (\overline e_{L\,p} \gamma_\mu e_{L\,r})  (i\phi_a \overleftrightarrow{\partial^\mu} \phi_b) & n_g^2  \binom{n_\phi}{2} & \Wcci{\phi l}{prab}{} -\frac{\overline g_Z^2}{M_Z^2}[Z_{e_L}]_{pr}[Z_\phi]_{ab}\\
& \opp{\phi u}{L} & (\overline u_{L\,p} \gamma_\mu u_{L\,r})  (i\phi_a \overleftrightarrow{\partial^\mu} \phi_b) & n_g^2  \binom{n_\phi}{2}& \Wcci{\phi q}{prab}{} -\frac{\overline g_Z^2}{M_Z^2}[Z_{u_L}]_{pr}[Z_\phi]_{ab}\\
& \opp{\phi d}{L}  & (\overline d_{L\,p} \gamma_\mu d_{L\,r})  (i\phi_a \overleftrightarrow{\partial^\mu} \phi_b) & n_g^2  \binom{n_\phi}{2} & \Wcci{\phi q}{prab}{} -\frac{\overline g_Z^2}{M_Z^2}[Z_{d_L}]_{pr}[Z_\phi]_{ab}\\
& \opp{\phi e}{R}  & (\overline e_{R\,p} \gamma_\mu e_{R\,r})  (i\phi_a \overleftrightarrow{\partial^\mu} \phi_b) & n_g^2  \binom{n_\phi}{2} & \Wcci{\phi e}{prab}{} -\frac{\overline g_Z^2}{M_Z^2}[Z_{e_R}]_{pr}[Z_\phi]_{ab}\\
& \opp{\phi u}{R}  & (\overline u_{R\,p} \gamma_\mu u_{R\,r})  (i\phi_a \overleftrightarrow{\partial^\mu} \phi_b) &  n_g^2  \binom{n_\phi}{2} & \Wcci{\phi u}{prab}{} -\frac{\overline g_Z^2}{M_Z^2}[Z_{u_R}]_{pr}[Z_\phi]_{ab}\\
& \opp{\phi d}{R}  & (\overline d_{R\,p} \gamma_\mu d_{R\,r})  (i\phi_a \overleftrightarrow{\partial^\mu} \phi_b) &  n_g^2  \binom{n_\phi}{2}   & \Wcci{\phi d}{prab}{} -\frac{\overline g_Z^2}{M_Z^2}[Z_{d_R}]_{pr}[Z_\phi]_{ab}\\
& \opp{eX\phi}{} & (\overline e_{L\,p} \sigma_{\mu \nu} e_{R\,r}) X_{a}^{\mu \nu} \phi_b + \text{h.c.} & n_g^2  n_\phi  n_X+ \text{h.c.}  & 0 \\
& \opp{uX\phi}{} & (\overline u_{L\,p} \sigma_{\mu \nu} u_{R\,r}) X_{a}^{\mu \nu} \phi_b + \text{h.c.} & n_g^2  n_\phi n_X + \text{h.c.}  & 0 \\
& \opp{dX\phi}{} & (\overline d_{L\,p} \sigma_{\mu \nu} d_{R\,r}) X_{a}^{\mu \nu} \phi_b + \text{h.c.} & n_g^2 n_\phi n_X  + \text{h.c.} & 0 \\
\bottomrule
\end{array}
\end{align*}
\caption{Dimension-six $\Delta B = \Delta L = 0$ operators in DLEFT, part 2. The first column gives the dimensions of the SM and DM part of the operator, the fourth column is the number of operators,  and the fifth column is the additional matching contribution  at the EW scale.}
\label{tab:DLEFTdim62}
\end{table}
%%%
% --- END TABLE
%%%

%%%
% --- START TABLE
%%%
\begin{table}[H]
\renewcommand{\arraystretch}{1.5}
\small
\begin{align*}
\begin{array}[t]{c|c|c|c|c}
\multicolumn{2}{c}{\textbf{DLEFT: dimension 6}} & \multicolumn{3}{c}{\boldsymbol{\Delta B = \Delta L = 0\,}} \\
\toprule
 (\text{d}_\text{SM}, \text{d}_{\text{DM}})& \text{Name} & \text{Operator} &  \text{Number} &  \text{Matching}  \\
\midrule\midrule
\multirow{16}{*}{(3,3)}
& \op{\nu\chi}{V}{LR} & (\overline \nu_{L\,p}\gamma_\mu \nu_{L\,r})(\overline \chi_a \gamma^\mu \chi_b) & n_g^2 n_\chi^2 \ & \Wcci{l\chi}{prab}{}-\frac{\overline g_Z^2}{M_Z^2}[Z_\nu]_{pr} [Z_\chi]_{ab} \\
& \op{e\chi}{V}{LR} & (\overline e_{L\,p}\gamma_\mu e_{L\,r})(\overline \chi_a \gamma^\mu \chi_b) & n_g^2 n_\chi^2 \ & \Wcci{l\chi}{prab}{} -\frac{\overline g_Z^2}{M_Z^2} [Z_{e_L}]_{pr} [Z_\chi]_{ab} \\
& \op{u\chi}{V}{LR} &  (\overline u_{L\,p}\gamma_\mu u_{L\,r})(\overline \chi_a \gamma^\mu \chi_b) & n_g^2 n_\chi^2 \ & \Wcci{q\chi}{prab}{} -\frac{\overline g_Z^2}{M_Z^2} [Z_{u_L}]_{pr} [Z_\chi]_{ab} \\
& \op{d\chi}{V}{LR} & (\overline d_{L\,p}\gamma_\mu d_{L\,r})(\overline \chi_a \gamma^\mu \chi_b) & n_g^2 n_\chi^2 & \Wcci{q\chi}{prab}{} -\frac{\overline g_Z^2}{M_Z^2} [Z_{d_L}]_{pr} [Z_\chi]_{ab} \\
& \op{e\chi}{V}{RR} &  (\overline e_{R\,p}\gamma_\mu e_{R\,r})(\overline \chi_a \gamma^\mu \chi_b) & n_g^2 n_\chi^2 \ & \Wcci{e\chi}{prab}{} -\frac{\overline g_Z^2}{M_Z^2} [Z_{e_R}]_{pr} [Z_\chi]_{ab} \\
& \op{u\chi}{V}{RR} &  (\overline u_{R\,p}\gamma_\mu u_{R\,r})(\overline \chi_a \gamma^\mu \chi_b) & n_g^2 n_\chi^2 \ & \Wcci{u\chi}{prab}{} -\frac{\overline g_Z^2}{M_Z^2} [Z_{u_R}]_{pr} [Z_\chi]_{ab} \\
& \op{d\chi}{V}{RR} & (\overline d_{R\,p}\gamma_\mu d_{R\,r})(\overline \chi_a \gamma^\mu \chi_b) & n_g^2 n_\chi^2 \ & \Wcci{d\chi}{prab}{} -\frac{\overline g_Z^2}{M_Z^2} [Z_{d_R}]_{pr} [Z_\chi]_{ab} \\
& \op{e\chi}{S}{RR} & (\overline e_{L\,p} e_{R\,r})(\chi_a^T C\chi_b) + \text{h.c.} & n_g^2 \binom{n_\chi+1}{2}   + \text{h.c.}  & 0 \\
& \op{u\chi}{S}{RR} &  (\overline u_{L\,p} u_{R\,r})(\chi_a^T C\chi_b) + \text{h.c.} & n_g^2  \binom{n_\chi+1}{2}   + \text{h.c.}  & 0 \\
& \op{d\chi}{S}{RR} & (\overline d_{L\,p} d_{R\,r})(\chi_a^T C\chi_b) + \text{h.c.} & n_g^2  \binom{n_\chi+1}{2}   + \text{h.c.}  & 0 \\
& \op{e\chi}{T}{RR} & (\overline e_{L\,p}  \sigma^{\mu \nu} e_{R\,r})(\chi_a^T C  \sigma_{\mu \nu}  \chi_b) + \text{h.c.} & n_g^2 \binom{n_\chi}{2}  + \text{h.c.}  & 0 \\
& \op{u\chi}{T}{RR} &  (\overline u_{L\,p}  \sigma^{\mu \nu}  u_{R\,r})(\chi_a^T C  \sigma_{\mu \nu}  \chi_b) + \text{h.c.} & n_g^2 \binom{n_\chi}{2}   + \text{h.c.}  & 0 \\
& \op{d\chi}{T}{RR} & (\overline d_{L\,p} \sigma^{\mu \nu} d_{R\,r})(\chi_a^T C  \sigma_{\mu \nu}  \chi_b) + \text{h.c.} & n_g^2 \binom{n_\chi}{2}  + \text{h.c.}  & 0 \\
& \op{e\chi}{S}{LR} & (\overline e_{R\,p} e_{L\,r})(\chi_a^T C\chi_b) + \text{h.c.} & n_g^2  \binom{n_\chi+1}{2}   + \text{h.c.}   &  0 \\
& \op{u\chi}{S}{LR} & (\overline u_{R\,p} u_{L\,r})(\chi_a^T C\chi_b) + \text{h.c.} & n_g^2  \binom{n_\chi+1}{2}   + \text{h.c.}   & 0 \\
& \op{d\chi}{S}{LR} & (\overline d_{R\,p} d_{L\,r})(\chi_a^T C\chi_b) + \text{h.c.} &  n_g^2   \binom{n_\chi+1}{2}   + \text{h.c.}  & 0 \\
\bottomrule
\end{array}
\end{align*}
\caption{Dimension-six $\Delta B = \Delta L = 0$ operators in DLEFT, part 3. The first column gives the dimensions of the SM and DM part of the operator, the fourth column is the number of operators,  and the fifth column is the additional matching contribution  at the EW scale.}
\label{tab:DLEFTdim63}
\end{table}
%%%
% --- END TABLE
%%%

%%%
% --- START TABLE
%%%
\begin{table}[H]
\renewcommand{\arraystretch}{1.5}
\small
\begin{align*}
\begin{array}[t]{c|c|c|c|c}
\multicolumn{2}{c}{\textbf{DLEFT: dimension 6}} & \multicolumn{3}{c}{\boldsymbol{\Delta B = \Delta L = 0\,}} \\
\toprule
 (\text{d}_\text{SM}, \text{d}_{\text{DM}})& \text{Name} & \text{Operator} &  \text{Number} &  \text{Matching}  \\
\midrule\midrule
\multirow{4}{*}{(4,2)} & \opp{F\phi^2}{} & F_{\mu \nu}F^{\mu \nu} \phi_a\phi_b  &  \binom{n_\phi+1}{2}  & \overline c^2\, \Wcci{B\phi^2}{ab}{}+\overline s^2\, \Wcci{W\phi^2}{ab}{} \\
& \opp{\widetilde F\phi^2}{} & \widetilde F_{\mu \nu} F^{\mu \nu} \phi_a\phi_b &  \binom{n_\phi+1}{2}  & \overline c^2\, \Wcci{\widetilde B\phi^2}{ab}{}+\overline s^2\, \Wcci{\widetilde W\phi^2}{ab}{} \\
& \opp{G\phi^2}{} & G^A_{\mu \nu}G^{A\,\mu \nu} \phi_a\phi_b &  \binom{n_\phi+1}{2}   & \Wcci{G\phi^2}{ab}{} \\
& \opp{\widetilde G\phi^2}{} &  \widetilde G^A_{\mu \nu} G^{A\,\mu \nu} \phi_a\phi_b & \binom{n_\phi+1}{2}  & \Wcci{\widetilde G\phi^2}{ab}{} \\
\hline
\multirow{5}{*}{(5,1)} & \opp{eF\phi}{} & (\overline e_{L\,p} \sigma^{\mu \nu} e_{R\,r}) F_{\mu \nu} \phi_a + \text{h.c.} & n_g^2 n_\phi  + \text{h.c.} & 0 \\
& \opp{uF\phi}{} & (\overline u_{L\,p} \sigma^{\mu \nu} u_{R\,r}) F_{\mu \nu} \phi_a + \text{h.c.} & n_g^2 n_\phi   + \text{h.c.}& 0 \\
& \opp{dF\phi}{} & (\overline d_{L\,p} \sigma^{\mu \nu} d_{R\,r}) F_{\mu \nu} \phi_a + \text{h.c.} & n_g^2 n_\phi   + \text{h.c.}& 0 \\
& \opp{uG\phi}{} &  (\overline u_{L\,p} \sigma^{\mu \nu}T^A u_{R\,r}) G^A_{\mu \nu} \phi_a + \text{h.c.} & n_g^2 n_\phi  + \text{h.c.} & 0 \\
& \opp{dG\phi}{} & (\overline d_{L\,p} \sigma^{\mu \nu}T^A d_{R\,r}) G^A_{\mu \nu} \phi_a + \text{h.c.} & n_g^2 n_\phi  + \text{h.c.} & 0 \\
\bottomrule
\end{array}
\end{align*}
\caption{Dimension-six $\Delta B = \Delta L = 0$ operators in DLEFT, part 4. The first column gives the dimensions of the SM and DM part of the operator, the fourth column is the number of operators,  and the fifth column is the additional matching contribution  at the EW scale.}
\label{tab:DLEFTdim64}
\end{table}
%%%
% --- END TABLE
%%%

%%%
% --- START TABLE
%%%
\begin{table}[H]
\renewcommand{\arraystretch}{1.5}
\small
\begin{align*}
\begin{array}[t]{c|c|c|c|c}
\multicolumn{2}{c}{\textbf{DLEFT: dimension 3}} & \multicolumn{3}{c}{\boldsymbol{\Delta B=0,\ \Delta L =1 + \text{h.c.}\,}} \\
\toprule
 (\text{d}_\text{SM}, \text{d}_{\text{DM}})& \text{Name} & \text{Operator} &  \text{Number} &  \text{Matching}  \\
\midrule\midrule
(3/2,3/2) & \opp{\chi\nu}{} & ( \overline \chi_a \nu_{L\,p} ) & n_g n_\chi & -\frac{v_T}{\sqrt{2}}\Wcci{H\chi l}{ap}{} -\frac{v_T^3}{2\sqrt{2}}\Wcci{H^3\chi l}{ap}{}\\
\bottomrule
\end{array}
\end{align*}
\caption{Dimension-three $\Delta B=0$, $\Delta L = 1$ operator in DLEFT. The first column gives the dimensions of the SM and DM part of the operator, the fourth column is the number of operators,  and the fifth column is the additional matching contribution  at the EW scale. }
\setlength{\belowcaptionskip}{-1cm}
%\caption{Dimension-three $\Delta L = 1$ operator in the DLEFT. }
\label{tab:DLEFTDL1dim3}
\end{table}
%%%
% --- END TABLE
%%%

%%%
% --- START TABLE
%%%
\begin{table}[H]
\renewcommand{\arraystretch}{1.5}
\small
\begin{align*}
\begin{array}[t]{c|c|c|c|c}
\multicolumn{2}{c}{\textbf{DLEFT: dimension 4}} & \multicolumn{3}{c}{\boldsymbol{\Delta B = 0,\ \Delta L =1 + \text{h.c.}\,}} \\
\toprule
 (\text{d}_\text{SM}, \text{d}_{\text{DM}})& \text{Name} & \text{Operator} &  \text{Number} &  \text{Matching}  \\
\midrule\midrule
(3/2,5/2) & \opp{\chi\nu\phi}{} & ( \overline \chi_a \nu_{L\,p} ) \phi_b & n_g n_\phi n_\chi  & -\frac{v_T}{\sqrt{2}}\Wcci{H\chi l\phi}{apb}{} \\
\bottomrule
\end{array}
\end{align*}
\caption{Dimension-four $\Delta B=0$, $\Delta L = 1$ operator in DLEFT. The first column gives the dimensions of the SM and DM part of the operator, the fourth column is the number of operators,  and the fifth column is the additional matching contribution  at the EW scale. }
\label{tab:DLEFTDL1dim4}
\end{table}
%%%
% --- END TABLE
%%%

%%%
% --- START TABLE
%%%
\begin{table}[H]
\renewcommand{\arraystretch}{1.5}
\small
\begin{align*}
\begin{array}[t]{c|c|c|c|c}
\multicolumn{2}{c}{\textbf{DLEFT: dimension 5}} & \multicolumn{3}{c}{\boldsymbol{\Delta B = 0,\ \Delta L =1 + \text{h.c.}\,}} \\
\toprule
 (\text{d}_\text{SM}, \text{d}_{\text{DM}})& \text{Name} & \text{Operator} &  \text{Number} &  \text{Matching}  \\
\midrule\midrule
\multirow{2}{*}{(3/2,7/2)} & \opp{\chi\nu\phi^2}{} & ( \overline \chi_a\nu_{L\,p} )\phi_b\phi_c & n_g \binom{n_\phi+1}{2}  n_\chi  & -\frac{v_T}{\sqrt{2}}\Wcci{H\chi l\phi^2}{apbc}{} \\
& \opp{\chi\nu X}{} & ( \overline \chi_a \sigma^{\mu \nu} \nu_{L\,p}) X_{b\,\mu \nu} & n_g  n_\chi  n_X & -\frac{v_T}{\sqrt{2}}\Wcci{H\chi lX}{apb}{} \\
\hline
(7/2,3/2) & \opp{F\chi\nu}{} &  F_{\mu \nu}( \overline \chi_a \sigma^{\mu \nu} \nu_{L\,p})  & n_g n_\chi & \frac{v_T}{\sqrt{2}}\left(\overline c\,\Wcci{HB\chi l}{ap}{}-\overline s\,\Wcci{HW\chi l}{ap}{}\right) \\
\bottomrule
\end{array}
\end{align*}
\caption{Dimension-five $\Delta B=0$, $\Delta L = 1$ operators in DLEFT. The first column gives the dimensions of the SM and DM part of the operator, the fourth column is the number of operators,  and the fifth column is the additional matching contribution  at the EW scale. }
\label{tab:DLEFTDL1dim5}
\end{table}
%%%
% --- END TABLE
%%%

%%%
% --- START TABLE
%%%
\begin{table}[H]
\renewcommand{\arraystretch}{1.5}
\small
\begin{align*}
\begin{array}[t]{c|c|c|c|c}
\multicolumn{2}{c}{\textbf{DLEFT: dimension 6}} & \multicolumn{3}{c}{\boldsymbol{\Delta B = 0,\ \Delta L =1 + \text{h.c.}\,}} \\
\toprule
 (\text{d}_\text{SM}, \text{d}_{\text{DM}})& \text{Name} & \text{Operator} &  \text{Number} &  \text{Matching}  \\
\midrule\midrule
\multirow{5}{*}{(3/2, 9/2)} & \opp{\chi\nu\phi^3}{} & (\overline \chi_a \nu_{L\,p} )\phi_b\phi_c\phi_d & n_g   \binom{n_\phi+2}{3} n_\chi & 0 \\
& \opp{\chi\nu X\phi}{} &  ( \overline \chi_a \sigma^{\mu \nu} \nu_{L\,p}) X_{b\,\mu \nu}\phi_c & n_g n_\phi n_\chi n_X  & 0 \\
& \opp{\phi\chi\nu}{L} & (\chi^T_a C \gamma_\mu \nu_{L\,p})  (i\phi_b \overleftrightarrow{\partial^\mu} \phi_c) & n_g  \binom{n_\phi}{2} n_\chi & 0 \\
&  \op{\chi\nu\chi}{S}{LR} & ( \overline \chi_a \nu_{L\,p} )( \chi_b^T C  \chi_c)  & n_g  n_\chi \binom{n_\chi+1}{2}  & 0 \\
& \op{\chi\nu\overline\chi}{S}{LL} & ( \overline \chi_a \nu_{L\,p} )(\overline \chi_b C \overline \chi_c^T)    &  \frac13  n_g n_\chi (n_\chi^2 -1) & 0 \\
\hline
(7/2,5/2) & \opp{F\chi\nu\phi}{} &  F_{\mu \nu}( \overline \chi_a \sigma^{\mu \nu} \nu_{L\,p}) \phi_b & n_g  n_\phi n_\chi & 0 \\
\hline
\multirow{5}{*}{(9/2, 3/2)}  & \op{\nu\chi\nu}{V}{LL} & (\overline \nu_{L\,p} \gamma^\mu \nu_{L\,r}) (\chi^T_a C \gamma_\mu \nu_{L\,s}) & n_g \binom{n_g+1}{2}  n_\chi  &  0\\
& \op{e\chi\nu}{V}{LL} & (\overline e_{L\,p} \gamma^\mu e_{L\,r}) (\chi^T_a C \gamma_\mu \nu_{L\,s}) & n_g^3 n_\chi  & 0 \\
& \op{u\chi\nu}{V}{LL} & (\overline u_{L\,p} \gamma^\mu u_{L\,r}) (\chi^T_a C \gamma_\mu \nu_{L\,s}) & n_g^3 n_\chi  & 0 \\
& \op{d\chi\nu}{V}{LL} & (\overline d_{L\,p} \gamma^\mu d_{L\,r}) (\chi^T_a C \gamma_\mu \nu_{L\,s}) & n_g^3 n_\chi & 0 \\
& \op{du\chi e}{V}{LL} &  (\overline d_{L\,p}\gamma_\mu u_{L\,r}) (\chi^T_a C \gamma^\mu e_{L\,s})  & n_g^3 n_\chi & 0 \\
\bottomrule
\end{array}
\end{align*}
\caption{Dimension-six $\Delta B=0$, $\Delta L = 1$ operators in DLEFT, part 1. The first column gives the dimensions of the SM and DM part of the operator, the fourth column is the number of operators,  and the fifth column is the additional matching contribution  at the EW scale.}
\label{tab:DLEFTDL1dim61}
\end{table}
%%%
% --- END TABLE
%%%

%%%
% --- START TABLE
%%%
\begin{table}[H]
\renewcommand{\arraystretch}{1.5}
\small
\begin{align*}
\begin{array}[t]{c|c|c|c|c}
\multicolumn{2}{c}{\textbf{DLEFT: dimension 6}} & \multicolumn{3}{c}{\boldsymbol{\Delta B = 0,\ \Delta L =1 + \text{h.c.}\,}} \\
\toprule
 (\text{d}_\text{SM}, \text{d}_{\text{DM}})& \text{Name} & \text{Operator} &  \text{Number} &  \text{Matching}  \\
\midrule\midrule
\multirow{21}{*}{(9/2,3/2)} & \op{e\chi\nu}{V}{RL} & (\overline e_{R\,p} \gamma^\mu e_{R\,r})  (\chi^T_a C \gamma_\mu \nu_{L\,s}) & n_g^3 n_\chi  & 0 \\
& \op{u\chi\nu}{V}{RL} & (\overline u_{R\,p} \gamma^\mu u_{R\,r}) (\chi^T_a C \gamma_\mu \nu_{L\,s}) & n_g^3 n_\chi  & 0 \\
& \op{d\chi\nu}{V}{RL} & (\overline d_{R\,p} \gamma^\mu d_{R\,r}) (\chi^T_a C \gamma_\mu \nu_{L\,s})  & n_g^3 n_\chi & 0 \\
& \op{du\chi e}{V}{RL} & (\overline d_{R\,p}\gamma_\mu u_{R\,r})  (\chi^T_a C \gamma^\mu e_{L\,s})& n_g^3 n_\chi & 0 \\
& \op{du\chi e}{V}{LR} & (\overline d_{L\,p}\gamma^\mu u_{L\,r} )( \overline \chi_a \gamma_\mu e_{R\,s})  & n_g^3 n_\chi & -\frac{\overline g_2^2}{2M_W^2} [W_q]^*_{rp} [W_\chi]_{as} \\
& \op{du\chi e}{V}{RR} & (\overline d_{R\,p}\gamma_\mu u_{R\,r})(\overline \chi_a \gamma^\mu e_{R\,s}) & n_g^3 n_\chi & \Wcci{du\chi e}{pras}{} -\frac{\overline g_2^2}{2M_W^2} [W_R]_{rp}^* [W_\chi]_{as}\\
& \op{du\chi e}{S}{RR} & (\overline d_{L\,p} u_{R\,r})(\chi^T_a C e_{R\,s}) & n_g^3 n_\chi  & 0 \\
& \op{du\chi e}{T}{RR} & (\overline d_{L\,p}\sigma_{\mu \nu} u_{R\,r})(\chi^T_a C\sigma^{\mu \nu} e_{R\,s}) & n_g^3 n_\chi  & 0 \\
& \op{e\chi\nu}{S}{RL} & (\overline e_{L\,p} e_{R\,r})  ( \overline \chi_a \nu_{L\,s}) & n_g^3 n_\chi  &
\frac{\overline g_2^2}{M_W^2} [W_l]_{sp}^* [W_\chi]_{ar}\\
& \op{u\chi\nu}{S}{RL} & (\overline u_{L\,p} u_{R\,r}) ( \overline \chi_a \nu_{L\,s}) & n_g^3 n_\chi  & \Wcci{qu\chi l}{pras}{} \\
& \op{d\chi\nu}{S}{RL} & (\overline d_{L\,p} d_{R\,r})  ( \overline \chi_a \nu_{L\,s})  & n_g^3 n_\chi & 0 \\
& \op{du\chi e}{S}{RL} & (\overline d_{L\,p} u_{R\,r}) ( \overline \chi_a e_{L\,s}) & n_g^3 n_\chi & \Wcci{qu\chi l}{pras}{} \\
& \op{e\chi\nu}{S}{LL} &  (\overline e_{R\,p} e_{L\,r})  ( \overline \chi_a \nu_{L\,s})  &  n_g^3 n_\chi  &
-\frac12 \Wcci{le\chi}{srpa}{(1)}  +\frac12  \Wcci{le\chi}{rspa}{(1)}  - 6 \Wcci{le\chi}{srpa}{(3)}  -6 \Wcci{le\chi}{rspa}{(3)}    \\
& \op{u\chi\nu}{S}{LL} &  (\overline u_{R\,p} u_{L\,r}) ( \overline \chi_a \nu_{L\,s})& n_g^3 n_\chi  & 0 \\
& \op{d\chi\nu}{S}{LL} & (\overline d_{R\,p} d_{L\,r}) ( \overline \chi_a \nu_{L\,s})   & n_g^3 n_\chi & -\Wcci{dq\chi l}{pras}{(1)}  \\
& \op{du\chi e}{S}{LL} & (\overline d_{R\,p} u_{L\,r}) ( \overline \chi_a e_{L\,s}) & n_g^3 n_\chi & \Wcci{dq\chi l}{pras}{(1)} \\
& \op{e\chi\nu}{T}{LL} &  (\overline e_{R\,p} \sigma^{\mu \nu}e_{L\,r}) (  \overline \chi_a \sigma^{\mu \nu} \nu_{L\,s})  &  n_g^3 n_\chi  & \frac18 \Wcci{le\chi}{srpa}{(1)}  - \frac18  \Wcci{le\chi}{rspa}{(1)}  - \frac12 \Wcci{le\chi}{srpa}{(3)}  - \frac12 \Wcci{le\chi}{rspa}{(3)}     \\
& \op{u\chi\nu}{T}{LL} &  (\overline u_{R\,p} \sigma^{\mu \nu}u_{L\,r}) (  \overline \chi_a \sigma^{\mu \nu} \nu_{L\,s})  & n_g^3 n_\chi  & 0 \\
& \op{d\chi\nu}{T}{LL} &  (\overline d_{R\,p}\sigma^{\mu \nu}d_{L\,r})  (  \overline \chi_a \sigma^{\mu \nu} \nu_{L\,s}) & n_g^3 n_\chi & -\Wcci{dq\chi l}{pras}{(3)} \\
& \op{du\chi e}{T}{LL} & (\overline d_{R\,p} \sigma_{\mu \nu} u_{L\,r})(  \overline \chi_a \sigma^{\mu \nu}  e_{L\,s}) & n_g^3 n_\chi & \Wcci{dq\chi l}{pras}{(3)} \\
& \op{du\chi e}{S}{LR} & (\overline d_{R\,p} u_{L\,r}) (\chi^T_a C e_{R\,s}) & n_g^3 n_\chi & 0 \\
\bottomrule
\end{array}
\end{align*}
\caption{Dimension-six $\Delta B=0$, $\Delta L = 1$ operators in DLEFT, part 2. The first column gives the dimensions of the SM and DM part of the operator, the fourth column is the number of operators,  and the fifth column is the additional matching contribution  at the EW scale.}
\label{tab:DLEFTDL1dim62}
\end{table}
%%%
% --- END TABLE
%%%

%%%
% --- START TABLE
%%%
\begin{table}[H]
\renewcommand{\arraystretch}{1.5}
\small
\begin{align*}
\begin{array}[t]{c|c|c|c|c}
\multicolumn{2}{c}{\textbf{DLEFT: dimension 4}} & \multicolumn{3}{c}{\boldsymbol{\Delta B = 0,\ \Delta L =2 + \text{h.c.}\,}} \\
\toprule
 (\text{d}_\text{SM}, \text{d}_{\text{DM}})& \text{Name} & \text{Operator} &  \text{Number} &  \text{Matching}  \\
\midrule\midrule
(3,1) & \opp{\nu\phi}{} & (\nu_{L\,p}^T C \nu_{L\,r}) \phi_a & \binom{n_g+1}{2}  n_\phi  & \frac{v_T^2}{2}\Wcci{\nu\nu\phi}{pra}{} \\
\bottomrule
\end{array}
\end{align*}
\caption{Dimension-four $\Delta B=0$, $\Delta L =2$ operator in DLEFT. The first column gives the dimensions of the SM and DM part of the operator, the fourth column is the number of operators,  and the fifth column is the additional matching contribution  at the EW scale.}
\label{tab:DLEFTDL2dim4}
\end{table}
%%%
% --- END TABLE
%%%

%%%
% --- START TABLE
%%%
\begin{table}[H]
\renewcommand{\arraystretch}{1.5}
\small
\begin{align*}
\begin{array}[t]{c|c|c|c|c}
\multicolumn{2}{c}{\textbf{DLEFT: dimension 5}} & \multicolumn{3}{c}{\boldsymbol{\Delta B = 0,\ \Delta L =2 + \text{h.c.}\,}} \\
\toprule
 (\text{d}_\text{SM}, \text{d}_{\text{DM}})& \text{Name} & \text{Operator} &  \text{Number} &  \text{Matching}  \\
\midrule\midrule
\multirow{2}{*}{(3,2)}  & \opp{\nu\phi^2}{} & (\nu_{L\,p}^T C \nu_{L\,r}) \phi_a\phi_b &   \binom{n_g+1}{2}  \binom{n_\phi+1}{2}   & 0 \\
& \opp{\nu X}{} &  ( \nu_{L\,p}^T C\sigma_{\mu \nu} \nu_{L\,r}) X_{a}^{\mu \nu} & \binom{n_\chi}{2}  n_X  & 0 \\
\bottomrule
\end{array}
\end{align*}
\caption{Dimension-five $\Delta B=0$, $\Delta L =2$ operators in DLEFT. The first column gives the dimensions of the SM and DM part of the operator, the fourth column is the number of operators,  and the fifth column is the additional matching contribution  at the EW scale.}
\label{tab:DLEFTDL2dim5}
\end{table}
%%%
% --- END TABLE
%%%

\vspace{2cm}

%%%
% --- START TABLE
%%%
\begin{table}[H]
\renewcommand{\arraystretch}{1.5}
\small
\begin{align*}
\begin{array}[t]{c|c|c|c|c}
\multicolumn{2}{c}{\textbf{DLEFT: dimension 6}} & \multicolumn{3}{c}{\boldsymbol{\Delta B = 0,\ \Delta L =2 + \text{h.c.}\,}} \\
\toprule
 (\text{d}_\text{SM}, \text{d}_{\text{DM}})& \text{Name} & \text{Operator} &  \text{Number} &  \text{Matching}  \\
\midrule\midrule
\multirow{5}{*}{(3,3)}  & \opp{\nu\phi^3}{} & ( \nu_{L\,p}^T C \nu_{L\,r}) \phi_a\phi_b\phi_c &   \binom{n_g+1}{2}   \binom{n_\phi+2}{3}  & 0 \\
& \opp{\nu X\phi}{} & (\nu_{L\,p}^T C\sigma_{\mu \nu} \nu_{L\,r}) X_{a}^{\mu \nu} \phi_b & \binom{n_g}{2}   n_\phi  n_X & 0 \\
& \op{\nu\overline\chi}{S}{LL} & (\nu_{L\,p}^T C \nu_{L\,r}) (\overline \chi_a C \overline \chi_b^T) &  \binom{n_g+1}{2}   \binom{n_\chi+1}{2}   & 0 \\
& \op{\nu\overline\chi}{T}{LL} &  (\nu_{L\,p}^T C\sigma_{\mu \nu} \nu_{L\,r}) (\overline \chi_a\sigma^{\mu \nu}C\overline \chi_b^T)  &  \binom{n_g}{2}   \binom{n_\chi}{2}  & 0 \\
& \op{\nu\chi}{S}{LR} & (\nu_{L\,p}^T C \nu_{L\,r})(\chi_a^T C \chi_b) &   \binom{n_g+1}{2}   \binom{n_\chi+1}{2}   &  0\\
\hline
(5,1) & \opp{\nu F\phi}{} & (\nu_{L\,p}^T C\sigma^{\mu \nu} \nu_{L\,r}) F_{\mu \nu}\phi_a & \binom{n_g}{2}   n_\phi  & 0 \\
\bottomrule
\end{array}
\end{align*}
\caption{Dimension-six $\Delta B=0$, $\Delta L =2$ operators in DLEFT. The first column gives the dimensions of the SM and DM part of the operator, the fourth column is the number of operators,  and the fifth column is the additional matching contribution  at the EW scale.}
\label{tab:DLEFTDL2dim6}
\end{table}
%%%
% --- END TABLE
%%%

%%%
% --- START TABLE
%%%
\begin{table}[H]
\renewcommand{\arraystretch}{1.5}
\small
\begin{align*}
\begin{array}[t]{c|c|c|c|c}
\multicolumn{2}{c}{\textbf{DLEFT: dimension 6}} & \multicolumn{3}{c}{\boldsymbol{\Delta B = 0,\ \Delta L =3 + \text{h.c.}\,}} \\
\toprule
 (\text{d}_\text{SM}, \text{d}_{\text{DM}})& \text{Name} & \text{Operator} &  \text{Number} &  \text{Matching}  \\
\midrule\midrule
(9/2,5/2) & \op{\nu \chi\nu}{S}{LL} & (\nu_{L\,p}^T C \nu_{L\,r}) (\overline \chi_a \nu_{L\,s}) & \frac13 n_g (n_g^2-1) n_\chi  & 0 \\
\bottomrule
\end{array}
\end{align*}
\caption{Dimension-six $\Delta B=0$, $\Delta L =3$ operator in DLEFT. The first column gives the dimensions of the SM and DM part of the operator, the fourth column is the number of operators,  and the fifth column is the additional matching contribution  at the EW scale.}
\label{tab:DLEFTDL3}
\end{table}
%%%
% --- END TABLE
%%%

%%%
% --- START TABLE
%%%
\begin{table}[H]
\renewcommand{\arraystretch}{1.5}
\small
\begin{align*}
\begin{array}[t]{c|c|c|c|c}
\multicolumn{2}{c}{\textbf{DLEFT: dimension 6}} & \multicolumn{3}{c}{\boldsymbol{\Delta L = 0,\ \Delta B=1 + \text{h.c.}\,}} \\
\toprule
 (\text{d}_\text{SM}, \text{d}_{\text{DM}})& \text{Name} & \text{Operator} &  \text{Number} &  \text{Matching}  \\
\midrule\midrule
\multirow{8}{*}{(9/2,3/2)} & \op{udd}{S}{LR} & \epsilon^{\alpha\beta\gamma}(u_{L\,p}^{\alpha\,T}Cd_{L\,r}^\beta)(\chi_a^T C d_{R\,s}^{\gamma} ) & n_g^3 n_\chi &  \Wcci{qqd}{pras}{} +\Wcci{qqd}{rpas}{}  \\
& \op{ddu}{S}{LL} &  \epsilon^{\alpha\beta\gamma}(d_{L\,p}^{\alpha\,T}C d_{L\,r}^\beta)(\overline \chi_a u_{L\,s}^{\gamma } ) & n_g  \binom{n_g}{2}  n_\chi & 0 \\
& \op{ddu}{T}{LL} &  \epsilon^{\alpha\beta\gamma}(d_{L\,p}^{\alpha\,T}C \sigma_{\mu \nu}  d_{L\,r}^\beta)(\overline \chi_a  \sigma^{\mu \nu}  u_{L\,s}^{\gamma}  ) &n_g  \binom{n_g+1}{2}n_\chi & 0 \\
& \op{ddu}{S}{RL} &  \epsilon^{\alpha\beta\gamma}(d_{R\,p}^{\alpha\,T}C d_{R\,r}^\beta)( \overline \chi_a   u_{L\,s}^{\gamma }  ) & n_g  \binom{n_g}{2} n_\chi & 0 \\
& \op{dud}{S}{RL} & \epsilon^{\alpha\beta\gamma}(d_{R\,p}^{\alpha\,T}Cu_{R\,r}^\beta)(\overline \chi_a d_{L\,s}^{\gamma } ) & n_g^3 n_\chi  & 0 \\
& \op{ddu}{S}{LR} &  \epsilon^{\alpha\beta\gamma}(d_{L\,p}^{\alpha\,T}C d_{L\,r}^\beta)(\chi_a^T C  u_{R\,s}^{\gamma } ) & n_g  \binom{n_g}{2}  n_\chi & 0 \\
& \op{ddu}{S}{RR} &  \epsilon^{\alpha\beta\gamma}(d_{R\,p}^{\alpha\,T}C d_{R\,r}^\beta)(\chi_a^T C u_{R\,s}^{\gamma} ) & n_g  \binom{n_g}{2}  n_\chi & \Wcci{ddu}{pras}{(1)} \\
& \op{ddu}{T}{RR} &  \epsilon^{\alpha\beta\gamma}(d_{R\,p}^{\alpha\,T}C \sigma^{\mu \nu} d_{R\,r}^\beta)(\chi_a^T C \sigma_{\mu \nu}  u_{R\,s}^{\gamma } ) & n_g  \binom{n_g+1}{2}  n_\chi & \Wcci{ddu}{pras}{(3)} \\
\bottomrule
\end{array}
\end{align*}
\caption{Dimension-six $\Delta L=0$, $\Delta B=1$ operators in DLEFT. The first column gives the dimensions of the SM and DM part of the operator, the fourth column is the number of operators,  and the fifth column is the additional matching contribution  at the EW scale.}
\label{tab:DLEFTDB1}
\end{table}
%%%
% --- END TABLE
%%%

%%%%%%%%%%%%%%%%%%%%%%%%%%%%%%%%%%%%%%%%%%%%%%%
\bibliographystyle{JHEP}
\bibliography{refs}

\end{document}